\documentclass[aps,prl,twocolumn,showpacs,groupedaddress,amsmath,amssymb]{revtex4}

\usepackage{braket}

\usepackage{graphicx}
\usepackage{color}
\usepackage{comment}
\usepackage{bm}
\usepackage{latexsym}
\newcommand{\tp}[1]{\textcolor{black}{#1}}

\begin{document}

\title{Towards a Floquet-Network Theory of Non-Reciprocal Transport}
\author{Huanan Li, Tsampikos Kottos} 
\affiliation{Wave Transport in Complex Systems Lab, Department of Physics, Wesleyan University, Middletown, CT-06459, USA}
\author{Boris Shapiro}
\affiliation{Technion - Israel Institute of Technology, Technion City, Haifa 32000, Israel}
\date{\today}

\begin{abstract}
We develop a theoretical framework that lays out the fundamental rules under which a periodic (Floquet) driving scheme can induce non-reciprocal 
transport. \tp{Our approach utilizes an extended Hilbert space where a Floquet network with an extra (frequency) dimension naturally arises. The 
properties of this network (its on-site potential and the intersite couplings) are in one-to-one correspondence with the initial driving scheme. Its 
proper design allows for a control of the multipath scattering processes and the associated interferences. We harness this degree of freedom to 
realize driving schemes with narrow or broad-band non-reciprocal transport.}
\end{abstract}

\maketitle
{\it Introduction --} Reciprocity - the invariance of a system under the exchange of the emitter and the receiver - as well as its violation, 
belongs to the most basic notions in physics. From the fundamental perspective its existence is associated with the presence of time-
reversal symmetry \cite{LL60,C45} - an inherent property of many equations that dictate (classical or quantum) wave propagation 
\cite{H84,ST07,MEW13,FSHA15}. On the technological level, the realization of non-reciprocal devices, like isolators and circulators, is 
at the forefront of our efforts for the next generation of communication schemes, imaging, quantum information etc. 

It is, therefore, not surprising that over the years the physics and engineering community tried to identify ways to induce non-reciprocal 
classical (and quantum) wave transport. For example, in the electromagnetics framework, non-reciprocal transport (NRT) is typically 
achieved via magneto-optical effects, \tp{like Faraday rotation or circular dichroism \cite{LL60,H84,ST07,ZK97}}. On many occasions, however, 
this approach is not satisfactory due to the weak nature of magneto-optical effects; typical situations involve on-chip photonics where one 
is bounded by the small footprints of the devices and by the requirement for the use of weak magnetic biases \cite{D05}. Another example 
appears in the acoustic framework. In this case the magneto-acoustic effects \tp{(like magneto-elastic effects \cite{L05})} are even weaker than 
their optical counterparts and thus can not be used for phononic NRT. Thus, alternative schemes that produce one-way structures for a 
general wave flow have been placed high on the agenda of the research community. 

An alternative approach to NRT invokes non-linear effects \cite{RKGC10,LC11,BFBRCEK13,Yang14,SYF15,LYC09,LGTZ10,BTD11}. 
Although in many situations (like in acoustics), this approach is the only viable alternative, it nevertheless suffers from a number of deficiencies. 
The most prominent - at least from the engineering perspective - is the fact that such asymmetric transport occurs above a certain power 
threshold of the input signal. Moreover, it mainly relies on frequency conversion, which significantly alters the incident signal.

\begin{figure}
\begin{center}
\includegraphics[width=1\columnwidth,keepaspectratio,clip]{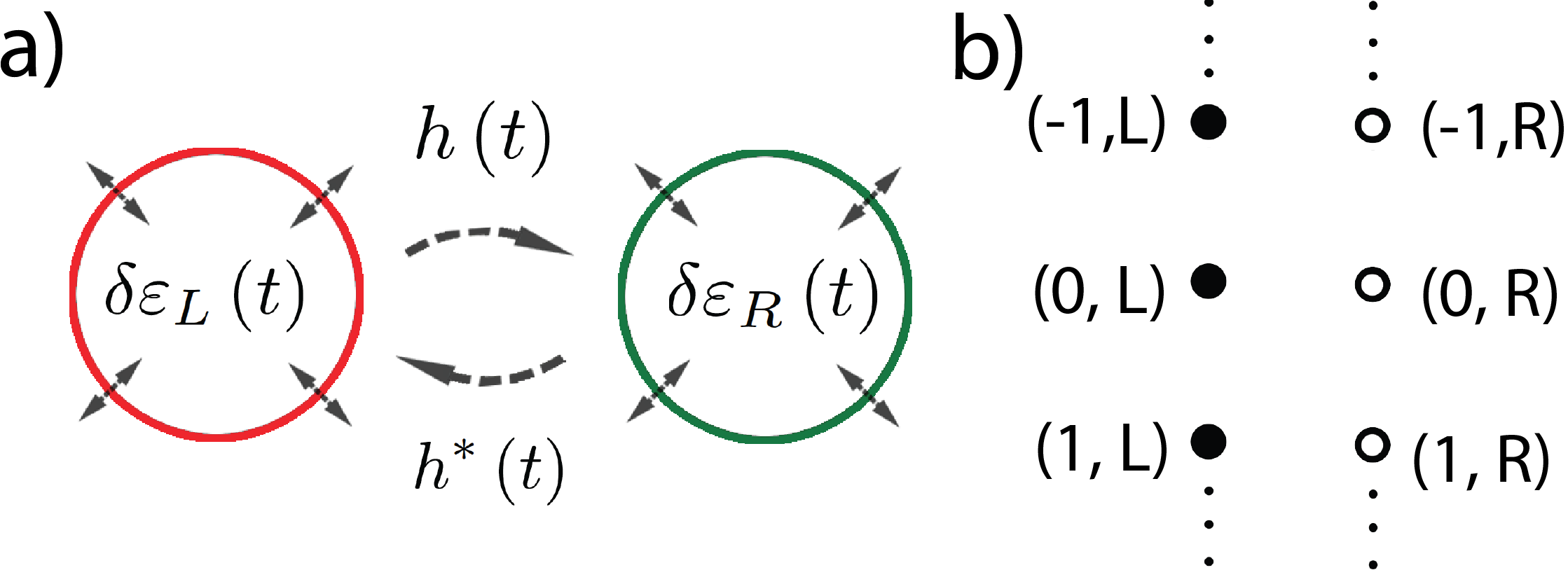}
\caption{(a) A system of two single-mode coupled cavities. The cavities can be passive or active i.e. consist of a loss or/and 
gain elements. A periodic modulation of the cavity (volume or constituent parameter) leads to a periodic modulation of 
the associated resonant frequencies. Their position can be also periodically modulated resulting in a modulation of their coupling 
strength. (b) The corresponding Floquet lattice. }
\label{fig1}
\end{center}
\end{figure}

A recent promising approach to NRT utilizes spatio-temporal modulations of the impedance profile of the medium using electrical, acoustic, 
mechanical, or optical biasing. \tp{Our capability to periodically modulate a system is known as Floquet engineering \cite{GD14,BAP15,E17} and 
its experimental verification can be found in various areas of physics \cite{SOTSELWS11,AALBPB13,MSKBK13,JMDLUGE14,RZPLPDNSS13}.} 
The approach resulted in a number of non-reciprocal schemes that capitalize on a variety of  physical mechanisms. Examples include non-reciprocal 
frequency conversion based on interband transitions \cite{YF09,LYFL12,Zhang}; azimuthal impedance modulations that create angular momentum-
bias in a ring-like geometry \cite{ESSA14,SCA13,FSSHA14}; or appropriate choices of spatial distribution of modulation phases that leads to 
effective magnetic fields \cite{FYF12,KA14,LEFF14}. Of course in on-chip photonics, many questions about the applicability of these methods 
can be raised. Some of these concerns, addresses our current capabilities to vary the electromagnetic properties of materials in space and time 
with sufficient depth and speed to break time reversal symmetry (TRS), the size of the non-reciprocal devices (much larger than the operating 
wavelength), etc.

Here we do not intend to address the technological aspects associated with the implementation of periodically driving systems for the 
realization of NRT. Instead, we address the following fundamental question: {\it Under which conditions a periodically modulated scheme can 
lead to NRT?} We provide a generalized framework which utilizes an extended Floquet space and allows us to identify the main ingredients of 
a driving scheme that leads to NRT. In this space the driving is described by a set of directed or/and directionless bonds which connects ``sites" 
in a Floquet network (lattice) whose on-site potential (resonant frequency) differs by a multiple of the driving frequency. Using a locator expansion 
approach, we show that the main mechanisms that lead to NRT are (a) the interferences between transmission paths consisting of directed and 
directionless bonds and (b) the radiative or Ohmic losses (or material gain) of the system. \tp{Furthermore, the proposed inverse engineering 
approach, provides an insight into {\it the control of NRT bandwidth (narrow vs broad-band) via a proper design of the Floquet network} (and thus 
of the corresponding periodic driving scheme)} \cite{note0}.  We demonstrate our theory using some paradigmatic examples which can be treated 
analytically \cite{supplement}. \tp{As a by-product of our study we have also analyzed (see \cite{supplement}) situations where non-reciprocal
up/down-conversion can be achieved.}

{\it  Floquet Lattices -} We consider two coupled single-mode resonators which are periodically driven, see Fig. \ref{fig1}a \tp{(for generalization 
to a multi-mode system see \cite{supplement})}. The driving can be associated with the coupling between the resonators or/and with the resonant 
frequencies of each of them. The coupled-mode equation for the driven system is determined by the time-periodic Hamiltonian $H\left(t+T\right)=H(t)$, 
\begin{align}
H\left(t\right)= & \left(\begin{array}{cc}
\varepsilon_{L}\left(t\right) & h\left(t\right)\\
h^{*}\left(t\right) & \varepsilon_{R}\left(t\right)
\end{array}\right),\label{eq:H_t}
\end{align}
where the coupling element $h(t)$ can be also, in general, complex, the superscript ``$*$" indicates complex conjugation, and the resonant 
angular frequency of each mode is $\varepsilon_{L}\left(t\right)$ and $\varepsilon_{R}\left(t\right)$. In general $\varepsilon_{L/R}(t)=\varepsilon^0_{L/R}
+\delta \varepsilon_{L/R}(t)$ where $\delta\varepsilon_{L/R}(t)$ is a real time-periodic variation and $\varepsilon_{L/R}^0$ can be complex due to material 
losses (or gain). 

The Hamiltonian in Eq. (\ref{eq:H_t}) is written in the ``site representation", where each of the resonators is represented by a site with the 
corresponding eigenstate $|\alpha\rangle$ ($\alpha =L,R$ for the left, right resonator) attached to the site. It is possible to replace the 
original time-dependent problem by a time-independent one, at the expense of an enlarged (infinite dimensional) Hilbert-Floquet space 
\cite{E17,Eck1}. The basis in this extended space is $|n,\alpha\rangle=|\alpha\rangle e^{-\imath n\omega t}$ with $\omega =2\pi/T$ , and the 
integer $n$ is running from $-\infty$ to $\infty$, see Fig.~\ref{fig1}b. A state vector $|\Psi\rangle$ in this basis is represented as $\Psi^T =(\cdots,\Psi_{n-1,L},
\Psi_{n-1,R},\Psi_{n,L},\Psi_{n,R},\Psi_{n+1,L},\Psi_{n+1,R},\cdots)$ and operators are represented by matrices with the characteristic 
block structure \cite{E17,Eck1}: the blocks are labeled by a pair $(n, m)$ while each block is a $2\times 2$ matrix in the ``internal" $(L, R)$
-space. For instance, the associated Hamiltonian \tp{$H_Q$} that describes the system of Eq. (\ref{eq:H_t}) in the extended space 
(sometimes called the ``quasienergy operator") is
$\left\langle n,\alpha\right| \tp{H_{Q}}\left|m,\beta \right\rangle =\left\langle n,\alpha\right| \tp{H_{Q}^0}\left|m,\beta \right\rangle-n\omega
\delta_{\alpha,\beta}\delta_{n,m}$ where $\left\langle n,\alpha\right| \tp{H_{Q}^0}\left|m,\beta \right\rangle=\frac{1}{T}\int_{0}^{T}\mathrm{dt}
e^{\imath\left(n-m\right)\omega t}H_{\alpha,\beta}\left(t\right)$. The diagonal elements of \tp{$H_Q$} are simply $(\cdots,\epsilon_L^0
+\omega,\epsilon_R^0+\omega,\epsilon_L^0,\epsilon_R^0,\epsilon_L^0-\omega,\epsilon_R^0-\omega,\cdots)$ \tp{and the off-diagonal 
part will be denoted as $V_Q$ where $V_Q=V_Q^{\dagger}$}. The latter is the connectivity (transition) 
matrix which determines the amplitudes for an excitation (i.e. the electric field) to perform transitions between the sites of the Floquet 
ladder (for an example see \cite{supplement}).

The transition amplitudes can be separated into purely real and imaginary elements associated with directionless and directed couplings in 
the Floquet lattice. If, for example, the driving function $h(t)$ in Eq. (\ref{eq:H_t}) contains a term $2h_n \sin(n\omega t)$, then an excitation 
at site $(s, L)$ in the Floquet lattice can penetrate to sites $(s\pm n, R)$ with a change $\mp n\omega$ in the frequency. The transition amplitude 
for this process is $\pm \imath h_n$ while it is $\mp \imath h_n$ in the opposite directions (directed coupling). The corresponding amplitude 
for $2h_n \cos(n\omega t)$ driving would be real, just $h_n$ (directionless coupling). Driving due to the on-site modulations $\delta \epsilon_{
L/R}(t)$ produces a different kind of couplings associated with up/down frequency conversions of the excitation at the specific left (L) or 
right (R) sites respectively. They are represented as vertical transitions $(n,L) \rightarrow (m,L)$ or $(n, R)\rightarrow (m, R)$ in the Floquet 
lattice. As in the case of $h(t)$-couplings, we have to distinguish between directed couplings (described by $\sin(n\omega t)$-terms) 
and directionless couplings (described by $\cos(n\omega t)$-terms). The transition rules (grammar), and their relation to the harmonics of 
the driving schemes are summarized in Fig. 2.

\begin{figure}
\begin{center}
\includegraphics[width=1\columnwidth,keepaspectratio,clip]{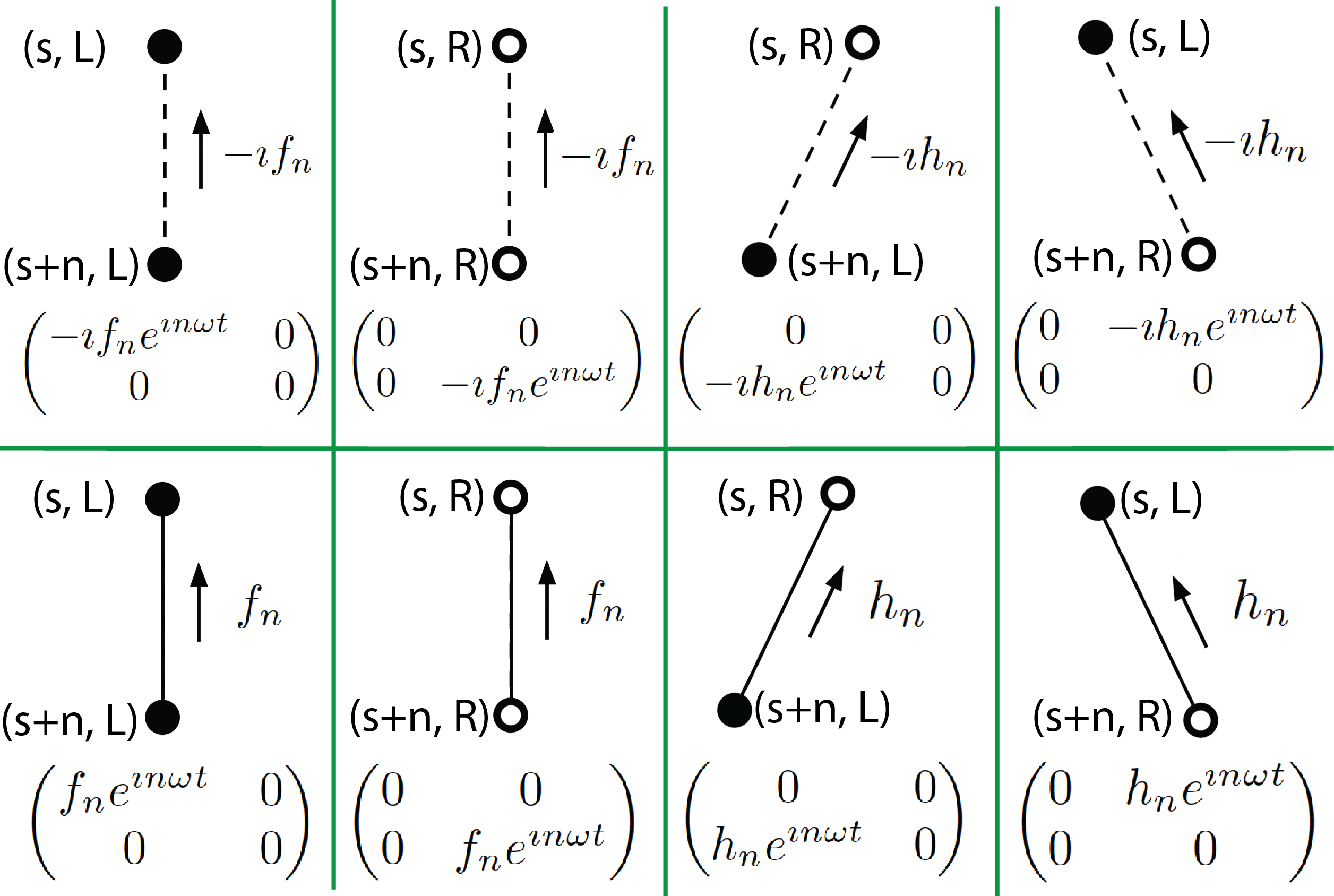}
 \caption{Filled (open) circles correspond to left (right) sites of the Floquet lattice. The index $s$ indicates the sites along the Floquet lattice while
the index $n$ indicates the distance between ladder steps. The magnitute of the bond strengths (Fourier components of the driving) are indicated as 
$f_n$ (for coupling between only L or only R sites of the Floquet lattice) and $h_n$ (for cross-coupling between L and R sites of the Floquet lattice). 
(First row): Grammar associated with imaginary driving amplitudes of Fourier components that correspond to directed bonds. (Second row): 
Grammar associated with real driving amplitudes of Fourier components that correspond to directionless bonds. In all cases, (upwards) arrows 
indicate transition to higher Fourier components. The associated driving schemes are also given in terms of matrices $H(t)$ as shown below each 
graph. The opposite transitions associated to complex conjugate amplitudes (downward arrows--not shown) are described by the Hermitian 
conjugate $H^{\dagger}(t)$.
}
\label{fig2}
\end{center}
\end{figure}

{\it Floquet Scattering-}  Let us now turn the discussion of NRT to its natural frame associated with a scattering problem. 
To this end, we couple each of the two resonators to identical leads which support plane waves with a dispersion relation $E=-2
\cos k$ (where the uniform coupling in the leads is set to be $-1$). For periodic driving, the Floquet theorem assures that when the incident 
wave with energy $E_{0}\in \left(-2,\,2\right]$ is approaching the time-periodic modulated target, it will scatter to infinite number of outgoing 
channels with energies $E_{n}=E_{0}+n\omega=-2\cos k_{n}$, where $n$ is integer and $k_{n}\in\left(0,\,\pi\right)$ for propagating channels and $\mathrm
{Im}k_{n}>0$ for evanescent channels. The corresponding Floquet $\mathcal{S}$ matrix, connects the outgoing to incoming waves (including 
evanescent channels), and can be written in the standard form used in the the case of time-independent scattering problems \cite{Data}
 (\tp{for derivation see \cite{supplement}})
\begin{eqnarray}
\mathcal{S} =  -1+2\imath W^{T}GW \label{eq: Floquet S} 
\end{eqnarray}
where the propagator $G$ is associated with the Hamiltonian \tp{$H_Q^0$} and can be evaluated using a locator diagrammatic 
expansion:
\begin{equation}
G\equiv -\frac{1}{D+\tp{H_{Q}^0}}=  G^{0}+G^{0}\tp{V_{Q}}G=G^{0}+G^0PG^0
\label{eq:G_dyson}
\end{equation}
Above $W,D$ are infinite-dimensional block-diagonal matrices with the $n-$th block being equal to $-\sqrt{\sin k_{n}}I_{2}$ and $e^{-\imath k_n} I_2$ 
respectfully ($I_2$ is the $2\times 2$ identity matrix). $G^{0}$ is the locator defined as
\begin{equation}
\label{locator}
G^{0}={\rm diag}(\cdots,g_{n-1}^{L}, g_{n-1}^{R},g_{n}^{L}, g_{n}^{R},g_{n+1}^{L}, g_{n+1}^{R},\cdots)
\end{equation}
with $g_{n}^{L/R}\equiv-1/\left(\varepsilon_{L/R}^{0}+e^{-\imath k_{n}}\right)$ incorporating the presence of channels. 
It is important to note that the propagating channels ($|E_n|\leq 2$) can introduce an imaginary part in $g_n^{L/R}$, and thus effectively 
play the role of loss in the system. Additionally the diagonal matrix $G^0$ (in the Floquet basis) might have complex elements inherited from 
$\epsilon_{L/R}^0$ due to material loss/gain. Finally, $P(n,L;m,R) \equiv 
\langle n,L|P|m,R\rangle= \langle n,L|\tp{V_Q}|m,R\rangle + \langle n,L|\tp{V_Q}G^0\tp{V_Q}|m,R\rangle +\cdots$.

Truncation of the Floquet $\mathcal{S}$ matrix to propagating channels produces the scattering matrix ${\tilde S}$, whose components
${\tilde S}(n,L;m,R)$ provide us with information about the NRT between left (L) and right (R) propagating channels with energies $E_n$ 
and $E_m$ respectively. Using Eq. (\ref{eq: Floquet S}) we get for the transmission probability amplitudes that ${\tilde S}\left(n,\,L;m,
\,R\right)=2\imath\sqrt{\sin k_{n}\sin k_{m}}G\left(n,\,L;m,\,R\right)$. This expression indicates that the NRT is dictated by the matrix elements 
of $G$ whose analysis will be given below. Specifically, we shall identify conditions under which $\left|\langle n,L|G|m,R\rangle\right|\neq 
\left|\langle m,R|G|n,L \rangle\right|$. This condition strongly depends on the special form of $G^0$ and \tp{$V_Q$}. 

Let us for example consider the matrix element $G(n,L;m,R)\equiv \langle n,L|G|m,R\rangle$. Using Eq. (\ref{eq:G_dyson}) with $G^0(n,L;m,R)=0$ 
(whenever $(n,L)\neq(m,R)$), we get
\begin{eqnarray}
\label{expansion2}
G(n,L;m,R)=G^0(n,L)G^0(m,R)\times P(n,L;m,R), 
\end{eqnarray}
where $G^0(n,L)\equiv G^0(n,L;n,L)$ indicates the diagonal term of the locator. Similarly one can calculate the matrix elements of $G(m,R;n,L)$
by exchanging $(n,L)\leftrightarrow (m,R)$ in Eq. (\ref{expansion2}). We find that these two transition probability amplitudes differ only by the 
term $P(n,L;m,R)$.

\tp{In the most general case $|P(n,L;m,R)|$ (and also $|P(m,R;n,L)|$) can be written as an infinite sum over all paths connecting the two Floquet 
sites $(n,L)\rightarrow (m,R)$. For a concise notation we have identified $\eta_l ({\tilde \eta}_l)$ as the probability amplitude of a directionless
(directional) path $l$. The probability amplitudes $\eta_l ({\tilde \eta}_l)$ involve multiplication of transition amplitudes (given by the off-diagonal 
matrix elements of $V_Q$) and the locators $g_n^{L/R}$ encountered at the path $l$, see Eq. (\ref{eq:G_dyson}). We re-organize the sum into 
two groups $P(n,L;m,R)=\eta (n,L;m,R) +{\tilde \eta}(n,L;m,R)$ where $\eta(n,L;m,R)=\sum_l\eta_l(n,L;m,R)$  and ${\tilde \eta}(n,L;m,R)=
\sum_l {\tilde \eta}_l(n,L;m,R)$ respectively. First we observe that forward and backward paths can be different whenever directed bonds are 
involved i.e ${\tilde \eta}_l(n,L;m,R)\neq {\tilde \eta}_l(m,R;n,L)$.  Another necessary condition is that the  imaginary parts of the locator elements 
have to be different from zero i.e. ${\cal I}m(g_n^{L/R})\neq 0$. Otherwise (even in the presence of directed bonds) the forward and backward 
paths will be complex conjugate of one another, leading to reciprocal transmissions. In the current frame the imaginary part of the locator originates 
from two possible sources: (a) the presence of material losses or gain in the resonators i.e. from the fact that ${\cal I}m(\epsilon_{L/R}^0)\neq 0$ 
and (b) due to the presence of other ``leaky" propagating channels coupled to the resonators.}

\tp{
When the directional group ${\tilde \eta}$ interfere strongly with the directionless one $\eta$, i.e. $|{\tilde \eta}|\sim |\eta|$, the NRT acquires 
a (local) maximum. In this case we have that $|P(n,L;m,R)|= |\eta| |1+e^{i\phi(n,L;m,R)}|$ and $|P(m,R;n,L)|=|\eta| |1+e^{i\phi(m,R;n,L)}|$. The relative 
phase $\Phi\equiv\phi\left(n,L;m,R\right)-\phi\left(m,R;n,L\right)$ is mainly controlled by the phase of the bonds and it is the primary source for 
NRT \cite{note1}. This phase depends weakly on the incident energy $E_0$ which mainly affects the locator(s) and consequently $|\eta|$ and $|{\tilde \eta}|$. 
Therefore any broad/narrow-band effects are traced to the behavior of the locator(s) inside an energy window $E_0^{\rm max}\pm \delta E$ 
($E_0^{\rm max}$ is the incident energy corresponding to the maximum NRT). }

\tp{The management of the NRT bandwidth $\delta E$ can be achieved by identifying the criteria for which the approximate relation $|{\tilde \eta}|
\sim |\eta|$ breaks down. Typically, the magnitude of the directionless group $\left|\eta\right|$ is of the order of the static coupling strength $h_{0}$ 
(DC term of $h(t)$). On the other hand, the magnitude of $\left|\tilde{\eta}\right|$ is controlled by two distinct 
mechanisms, associated with the dominant terms (paths ${\tilde h}_l$) within the sum ${\tilde \eta}$. To understand better their origin let 
us assume for simplicity that the dominant terms in ${\tilde \eta}$ contain two paths ${\tilde \eta}_{1}$ and ${\tilde \eta}_{2}$. Then we have 
$|{\tilde \eta}|\approx |{\tilde \eta}_{1}+{\tilde \eta}_{2}|$. This summation can be further factorized to two terms i.e. $|{\tilde \eta}|\approx B*L(E_0)$. 
The first term $B$ is energy independent and involves the product of bond-strengths. The second term $L(E_0)$ is associated with the product 
of locator(s) and it depends on the incident energy $E_0$. If $B\ll h_0$ then the approximate relation $|{\tilde \eta}|\sim |\eta|$ 
breaks down, apart from the situation where the incident energy $E_0$ is close to the locator resonance energy. In this case we have a narrow-band 
NRT. When $B\rightarrow h_0$ then the condition $|{\tilde \eta}|\sim |\eta|$ is satisfied when $L(E_0)\sim {\cal O}(1)$ i.e. when $E_0$ is 
away from the resonant energy of the locator and $L(E_0)$ is a smooth function of $E_0$. In this case the system supports a broad-band NRT.
It has to be clear that the dominant paths ${\tilde \eta}_{1,2}$ of the Floquet-network can be engineered in a way that will satisfy any of the conditions 
that we discuss above, thus producing broad/narrow-band NRT. Below we implement this strategy for three engineered Floquet-lattices that illustrate
NRT band-width management.}

{\it Examples of Non-Reciprocal Transport due to Floquet Driving --} We quantify the 
strength of NRT between the channels $E_{n}$ and $E_{m}$ of the left and right leads by introducing the non-reciprocity 
parameter $NR_{n\rightarrow m}$:
\begin{align}
NR_{n\rightarrow m}= & \frac{\left|{\tilde S}\left(m,\,R;n,\,L\right)\right|^{2}-\left|{\tilde S}\left(n,\,L;m,\,R\right)\right|^{2}}{\left|{\tilde S}
\left(m,\,R;n,\,L\right)\right|^{2}+\left|{\tilde S}\left(n,\,L;m,\,R\right)\right|^{2}}.\label{eq: NR}
\end{align}
Below, we consider the non-reciprocity strength $NR_{0\rightarrow0}$ between the channels $E_{0}$ of two leads for various driving schemes. 
Our choice of characterizing NRT via $NR_{0\rightarrow0}$ aims to demonstrate that the designed NRT schemes {\it do not depend on frequency 
conversion}.

\begin{figure}
\includegraphics[width=1\columnwidth,keepaspectratio,clip]{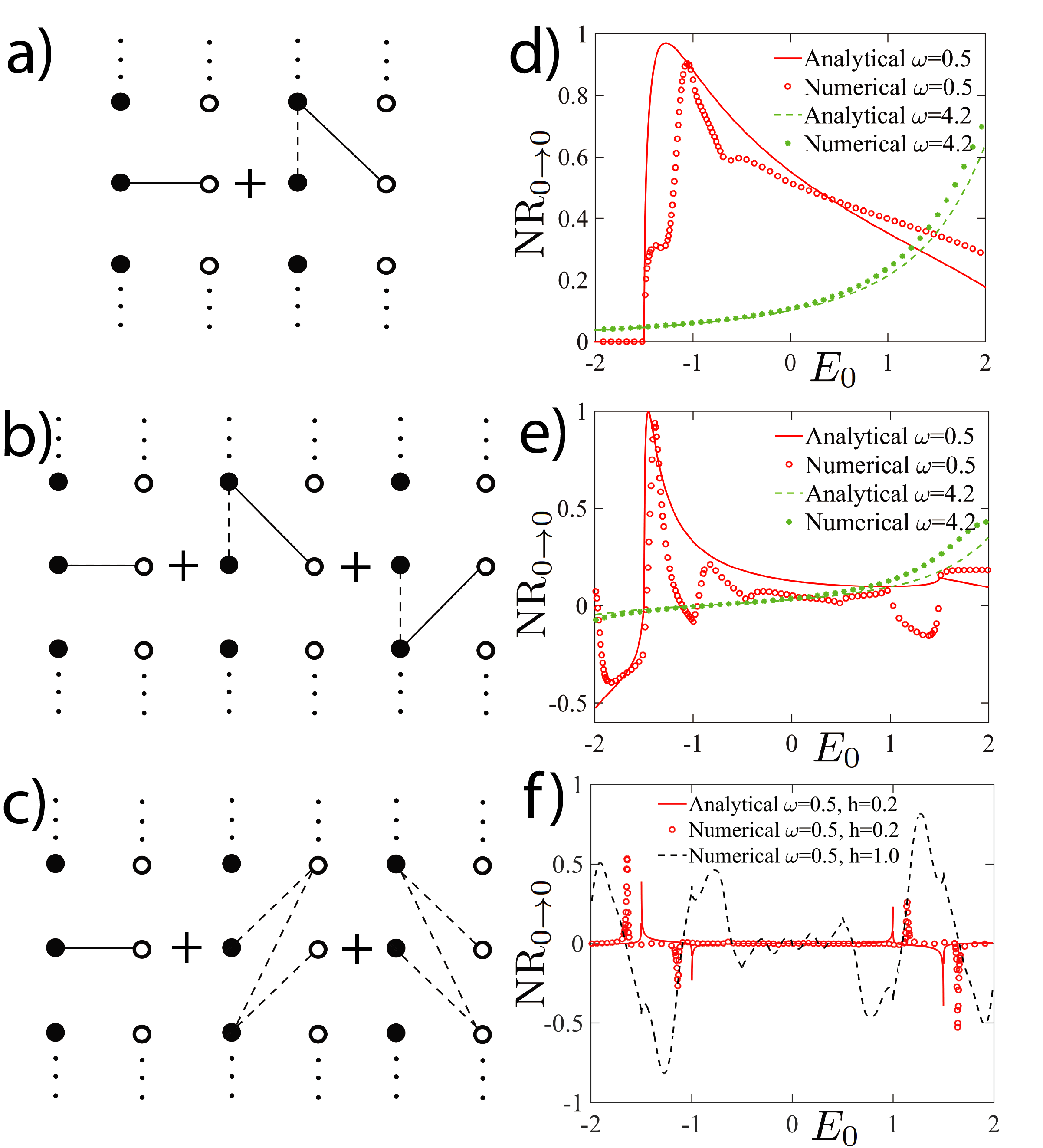}
\begin{center}
 \caption{Interference paths that have been considered in the evaluation of NR$_{0\rightarrow 0}$, associated with (a) Example 1, see 
Eq. (\ref{eq: path_approx}); (b) Example 2, see Eq. (\ref{eq: path_approx}) and (c) Example 3. Numerical and analytical evaluation of 
NR$_{0\rightarrow 0}$ versus incident frequency $E_0$ for $\varepsilon_{R}^{0}=0$, $f_{1}=h_{1}=0.5$ and (d) Example 1 with $h_0=-0.5$ 
and (e) Example 2 with $h_0=-1$. The simulations/calculations have been done for
two set of parameters $\omega=0.5$ and $\varepsilon_{L}^{0}=-1$, and for $\omega=4.2$ and $\varepsilon_{L}^{0}=-1-\imath$. (f) Numerical and
analytical
evaluation of NR$_{0\rightarrow 0}$ versus incident frequency $E_0$ for $\omega=0.5$, $\varepsilon_{L}^{0}=-1$, $\varepsilon_{R}^{0}=1$,
$h_{0}=-1$, $h_{1}=h_{2}=h=0.2$ (red line and symbols). \tp{The black dashed line indicates the case where a broad-band NRT is restored by using 
$h_1=h_2=h=1$}. The Floquet ladder in (a-c), is identical to Fig. 1b. \tp{The numerical convergence was tested by increasing the Floquet lattice.}
}
\label{fig3}
\end{center}
\end{figure}

{\it Example 1--}We start with a simple driving scheme associated with a periodically driven Hamiltonian $H(t)$ (see Eq. (\ref{eq:H_t})) where 
$h(t)=h_{0}+h_{1} e^{\imath\omega t}$, $\delta \epsilon_{L}(t)=2f_{1}\sin\omega t$ and $\delta \epsilon_R(t)=0$. In Fig. \ref{fig3}a we show the 
Floquet lattice and the associated interfering paths responsible for NRT -- one containing a directed bond and another one associated with a 
directionless bond. Under this approximation the Green's function from one direction is $G\left(0,\,R;0,\,L\right)=g_{0}^{L}g_{0}^{R}\left(h_{0}-
\imath f_{1}g_{-1}^{L}h_{1}\right)$ which is different from the Green function associated with the opposite direction $G\left(0,\,L;0, \,R\right)=
g_{0}^{L}g_{0}^{R}\left(h_{0}+h_{1}g_{-1}^{L}\imath f_{1}\right)$. Using Eqs. (\ref{eq: Floquet S},\ref{eq: NR}) we get for $NR_{0\rightarrow0}$ 
\begin{equation}
NR_{0\rightarrow0}\approx \frac{2h_{0}f_{1}h_{1}\mathrm{Im}\Delta}{h_{0}^{2}+f_{1}^{2}h_{1}^{2}\left|\Delta\right|^{2}}
\label{eq: path_approx}
\end{equation}
where $\Delta=g_{-1}^{L}$. Based on the approximated expression of Eq.~(\ref{eq: path_approx}), we consider two limiting cases associated 
with large and small driving frequencies, see Fig. \ref{fig3}d. In the former case, one needs to assume that $\mathrm{Im\varepsilon_{L}^{0}} 
\neq 0$ resulting in $\mathrm{Im}\Delta\neq0$ which guarantees that $NR_{0\rightarrow0}\neq0$. In the latter case of small frequencies, we 
shall typically have NRT irrespective of the form of $\varepsilon_{L}^{0}$. In this case $\mathrm{Im}\Delta\neq0$, due to the complex nature 
of $g_{-1}^{L}$ originating from the presence of the $E_{-1}$ propagating channel which plays the role of loss.

{\it Example 2--}A more realistic driving scheme can be obtained by a slight modification of the driving of Example 1. In this case $h(t)=h_{0}+
2h_{1}\cos\omega t$ while all other time-dependent parameters remain the same as previously. For this model, the nonreciprocity $NR_{0\rightarrow 
0}$ is still given by Eq.~(\ref{eq: path_approx}) but now with $\Delta=g_{-1}^{L}-g_{1}^{L}$. The associated paths that results in this expression 
are shown in Fig. \ref{fig3}b. When the incident wave has energy $E_{0}=0$ (middle of the band), the expression for $\Delta$ turns out to be 
$\mathrm{Im}\Delta\propto \mathrm{Re}\varepsilon_{L}^{0}$ irrespective of the driving frequency $\omega$. In this case, the condition 
$\mathrm{Re}\varepsilon_{L}^{0}\neq0$ is required in order to have $NR_{0\rightarrow0}\neq 0$, see Fig. \ref{fig3}e.

{\it Example 3--}Finally, we analyze a driving occuring via the coupling $h(t)$ between the resonators. 
The designed strategy for the driving function $h(t)$ \tp{and an analytical derivation of $NR_{0\rightarrow0}(E_0)$} is given in \cite{supplement}. 
Our engineered Floquet lattice produces (directional) interfering paths involving couplings up 
to next nearest neighbors, see Fig. \ref{fig3}c. The corresponding driving scheme has the form $h\left(t\right)=h_{0}+2h_{1}\sin\omega t+
2h_{2}\sin2\omega t$ with $\varepsilon_{L}\left(t\right)=\varepsilon_{L}^{0}$, and $\varepsilon_{R}\left(t\right)=\varepsilon_{R}^{0}$. The 
simulations for $NR_{0\rightarrow0}$ are shown in Fig. \ref{fig3}f. \tp{At the same figure we demonstrate
the NRT band-width management discussed above, within the same driving scheme. Specifically by modifying the driving amplitudes
$h_1,h_2$ we restored the condition $\eta\sim {\tilde \eta}$ and realized a broad-band NRT (see dashed black line in Fig. \ref{fig3}f and 
detail discussion related to Fig. S5 in \cite{supplement}).
}

{\it Acknowledgments --} 
This research was partially supported by an AFOSR grant No. FA 9550-10-1-0433, and by NSF grants EFMA-1641109 and DMR-1306984. BS 
acknowledges the hospitality of the Physics Department of Wesleyan University, where this work was performed. TK is grateful to I. Vitebskiy for 
many useful and illuminating discussions on non-reciprocal transport.

\onecolumngrid

\begin{center}
\textbf{\large Supplemental Materials}
\end{center}

\setcounter{equation}{0}
\setcounter{figure}{0}
\setcounter{table}{0}
\setcounter{page}{1}
\makeatletter
\renewcommand{\theequation}{S\arabic{equation}}
\renewcommand{\thefigure}{S\arabic{figure}}
\renewcommand{\bibnumfmt}[1]{[S#1]}
\renewcommand{\citenumfont}[1]{S#1}

\section{Example of a Floquet Lattice and of the associated driving}

Consider as an example a periodically modulated coupling with strength $h(t)=h_0+2h_1\sin(\omega t)+2 h_2 \cos(2\omega t)$. The term 
$h_0$ yields a coupling, not related to driving, between a pair of sites. Such horizontal couplings, for the pair $n=2$, is shown in Fig. S1 by 
a solid line (there are identical couplings for all other pairs, with any other value of $n$). Driving makes it possible to have transitions up and 
down the Floquet ladder. The term $2h_1\sin(\omega t)$ in $h(t)$ induces couplings between left and right resonators, with an amplitude 
$\mp i h_1$ and a change $\pm\omega$ in frequency. An example of such coupling is shown in the figure by the dashed line connecting site 
$(2,L)$ and $(1,R)$. The arrows indicate whether the transition is up or down the ladder i.e. if the frequency increases or decreases by 
$\omega$. Similarly, the $2 h_2 \cos(2\omega t)$ term describes couplings with amplitudes $h_2$ and changes $\pm 2\omega$ in frequency 
of the wave (the solid lines connecting site $(0,L)$ and $(-2,R)$ in the figure is an example of this kind of coupling). Note that the cosine 
driving, unlike the sine driving, results in a real amplitude $h_2$. Finally in the case of on-site driving, say $\delta \epsilon_L =2 f_1 \sin(\omega t)$,
a different type of couplings between the sites of the Floquet lattice are created. Specifically, such driving induces vertical couplings, like those 
shown in the Fig. S1 by the dashed line connecting site $(0,L)$ and $(1,L)$. 
\begin{figure}[h]
\includegraphics[width=0.35\columnwidth,keepaspectratio,clip]{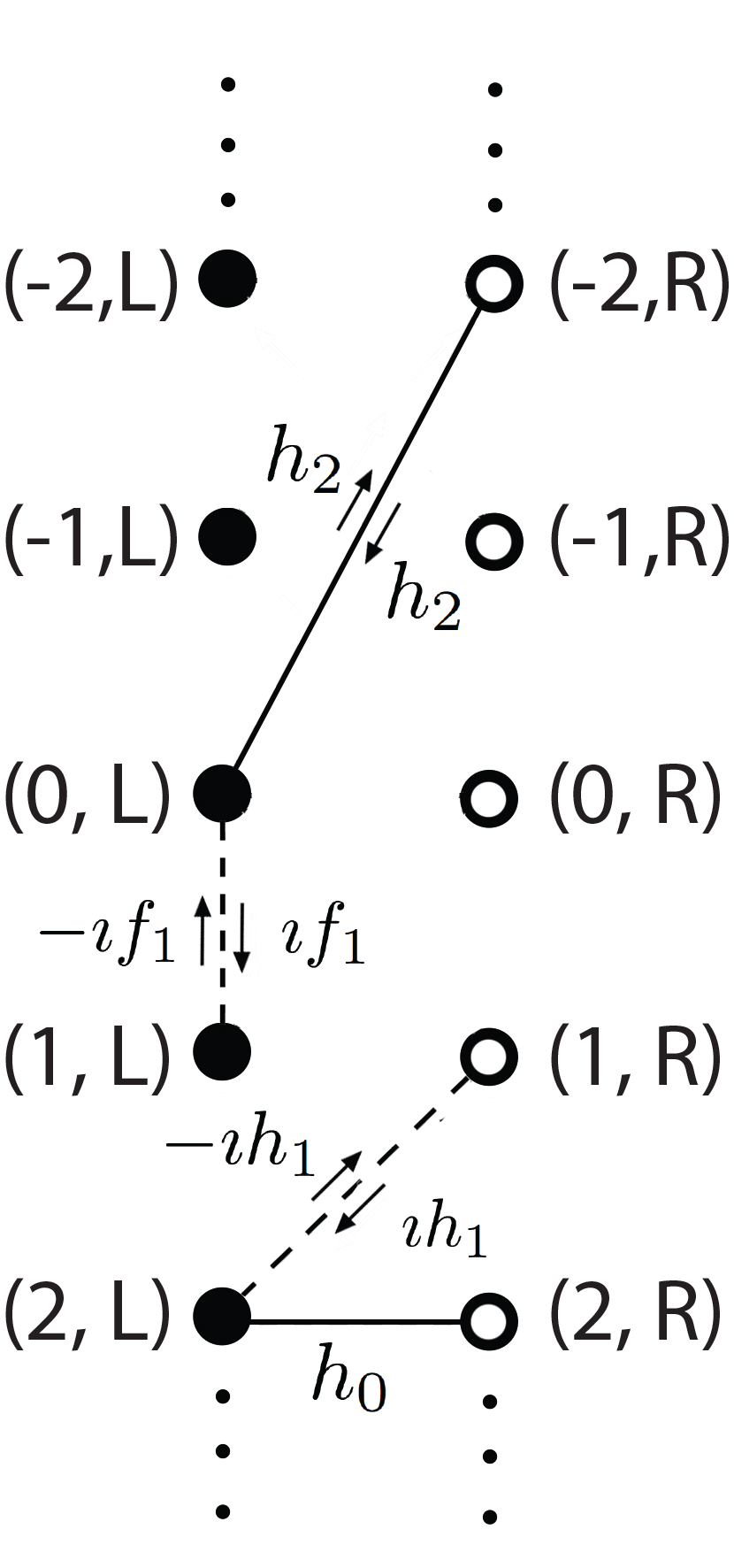}
\caption{ (color online) Example of a Floquet lattice with typical couplings due to the driving $h(t)=h_0+2h_1\sin(\omega t)+2 
h_2 \cos(2\omega t)$ and $\delta \epsilon_L =2 f_1 \sin(\omega t)$.
}
\label{figS1}
\end{figure}

\section{Derivation of Floquet $S$ Matrix}

A Floquet scattering state, consists of a scattering amplitude $\psi_{l}\left(t\right)$ at site $l$ of the scattering set-up (see the main text) 
which is given as 
\begin{align}
\psi_{l}\left(t\right)= & \begin{cases}
\sum_{n=-\infty}^{\infty}\left(L_{n}^{i}\frac{1}{\sqrt{\sin k_{n}}}e^{-\imath E_{n}t+\imath k_{n}l}+L_{n}^{o}\frac{1}{\sqrt{\sin k_{n}}}e^{-\imath 
E_{n}t-\imath k_{n}l}\right), & l\leqslant0\\
\sum_{n=-\infty}^{\infty}\left(R_{n}^{i}\frac{1}{\sqrt{\sin k_{n}}}e^{-\imath E_{n}t-\imath k_{n}l}+R_{n}^{o}\frac{1}{\sqrt{\sin k_{n}}}e^{-\imath 
E_{n}t+\imath k_{n}l}\right), & l\geqslant1
\end{cases}.
\label{eq: ansatz0}
\end{align}
Above $L_{n}^{i\left(o\right)}/R_{n}^{i\left(o\right)}$ represents the flux-normalized left/right propagating wave amplitude for an input (output) 
at channel $E_{n}=E_{0}+n\omega=-2\cos k_{n}$ where $E_{0}\in\left(-2,2\right]$, $n$ is an integer and $k_{n}\in\left(0,\pi\right)$ for propagating 
channels and $\mathrm{Im}k_{n}>0$ for evanescent channels.

The sites $l=0,1$ are special, since they label the position of the two coupled single-mode periodically driven resonators. The associated coupled
-mode equation which describes the temporal evolution of the field amplitude at these sites reads 
\begin{align}
\imath\frac{\mathrm{d}}{\mathrm{dt}}\left(\begin{array}{c}
\psi_{0}\\
\psi_{1}
\end{array}\right)= & H\left(t\right)\left(\begin{array}{c}
\psi_{0}\\
\psi_{1}
\end{array}\right)-\left(\begin{array}{c}
\psi_{-1}\\
\psi_{2}
\end{array}\right)\label{eq: CMeq}
\end{align}
where $H\left(t\right)=H\left(t+T\right)$ with $T=2\pi/\omega$. Combining Eqs.~(\ref{eq: ansatz0}) and (\ref{eq: CMeq}), we get 
\begin{align}
\tilde{\psi}_{L}= & \left[H\left(t\right)-\imath\frac{\mathrm{d}}{\mathrm{dt}}-E_{0}\right]\tilde{\psi}_{S},\label{eq: key}
\end{align}
where the subscripts $L(S)$ indicate the leads (system) and $\tilde{\psi}_{L/S}=\sum_{n}\tilde{\psi}_{L/S}^{\left(n\right)}e^{-\imath n\omega t}$
with $\tilde{\psi}_{L}^{\left(n\right)}=\frac{1}{\sqrt{\sin k_{n}}}\left(\begin{array}{c}
L_{n}^{i}e^{-\imath k_{n}}+L_{n}^{o}e^{\imath k_{n}}\\
R_{n}^{i}e^{-2\imath k_{n}}+R_{n}^{o}e^{2\imath k_{n}}
\end{array}\right)$ and $\tilde{\psi}_{S}^{\left(n\right)}=\frac{1}{\sqrt{\sin k_{n}}}\left(\begin{array}{c}
L_{n}^{i}+L_{n}^{o}\\
R_{n}^{i}e^{-\imath k_{n}}+R_{n}^{o}e^{\imath k_{n}}
\end{array}\right)$. In the extended Hilbert-Floquet space where a complete set of orthonormal basis is given by $\left|n,\alpha\right\rangle = 
\left|\alpha\right\rangle e^{-\imath n\omega t}$, Eq.~(\ref{eq: key}) can be easily converted into \cite{EA15}
\begin{align}
\tilde{\psi}_{L}^{F}= & \left(H_{Q}-E_{0}\right)\tilde{\psi}_{S}^{F},\label{eq: beforeS}
\end{align}
where $\left(\tilde{\psi}_{L/S}^{F}\right)^{T}\equiv\left(\cdots,\tilde{\psi}_{L/S}^{\left(n-1\right),T},\tilde{\psi}_{L/S}^{\left(n\right),T},
\tilde{\psi}_{L/S}^{\left(n+1\right),T},\cdots\right)$ and $\ensuremath{\left\langle n,\alpha\right| H_{Q}\left|m,\beta\right\rangle =
\left\langle n,\alpha\right|H_{Q}^{0}\left|m,\beta\right\rangle -n\omega\delta_{\alpha,\beta}\delta_{n,m}}$ with $\ensuremath{\left\langle n,
\alpha\right|H_{Q}^{0}\left|m,\beta\right\rangle =\frac{1}{T}\int_{0}^{T}\mathrm{dt}e^{\imath\left(n-m\right)\omega t}H_{\alpha,\beta}\left(t\right)}$.

A simple re-arrangement of Eq.~(\ref{eq: beforeS}) allows us to express the outgoing wave amplitudes in terms of the incoming ones. 
Specifically Eq.~(\ref{eq: beforeS}) can be rewritten in the form $\ensuremath{\varPsi^{o}=\mathcal{S}\varPsi^{i}}$,
where the input/output amplitudes are
\begin{align}
\left(\varPsi^{i/o}\right)^{T}\equiv & \left(\cdots,L_{n-1}^{i/o},R_{n-1}^{i/o}e^{\mp\imath k_{n-1}},L_{n}^{i/o},R_{n}^{i/o}e^{\mp\imath k_{n}},
L_{n+1}^{i/o},R_{n+1}^{i/o}e^{\mp\imath k_{n+1}},\cdots\right)
\label{eq: I_O}
\end{align}
and $\mathcal{S}$ is identified as the Floquet scattering matrix which can be written as in Eq. (2) of the main text.

\tp{
\section{Alternative derivation and generalization to multi-mode systems} }
\tp{
Here we present an alternative derivation for the Floquet $S$ matrix which allows for generalization to multi-mode systems. We consider 
a multi-mode system coupled to two leads. We assume that the system-lead coupling occurs via different modes. The system is subject to 
an arbitrary time-periodic driving. The Hamiltonian of the driven isolated system (i.e. in the absence of the leads) is denoted as 
$H\left(t\right)=H\left(t+T\right)$ where $T=2\pi/\omega$. For simplicity, we assume that the leads are described by a $1D$ tight-binding 
dispersion, $i.e.,$ $E=-2\cos k$. A general Floquet scattering state is specified by a scattering amplitude $\psi^{\alpha}\left(x_{\alpha},t\right)$
at site $x_{\alpha}$ of the lead $\alpha=L,R$ and is given as}
\tp{
\begin{align}
\psi^{\alpha}\left(x_{\alpha},t\right)= & \sum_{n=-\infty}^{\infty}\left(\alpha_{n}^{i}\frac{1}{\sqrt{\sin k_{n}}}e^{-\imath E_{n}t+\imath k_{n}
x_{\alpha}}+\alpha_{n}^{o}\frac{1}{\sqrt{\sin k_{n}}}e^{-\imath E_{n}t-\imath k_{n}x_{\alpha}}\right).\label{eq: ansatz}
\end{align}   }
\tp{
Above the tight-binding site $x_{\alpha}$ take integer values from $-\infty$ to $0$ where $x_{\alpha}=0$ indicates the site where the lead 
$\alpha$ is coupled with the system. The channels at each lead are indicated as $E_{n}=E_{0}+n\omega=-2\cos k_{n}$ ($n$ is an integer)
where $E_{0}\in\left(-2,2\right]$. In each lead, there exist both propagating $E_{n}\in\left(-2,2\right)$ and evanescent $E_{n}\notin\left(
-2,2\right)$ channels. The propagating channels of the lead $\alpha$, correspond to wavenumbers $k_{n}\in\left(0,\pi\right)$ where 
$\alpha_{n}^{i\left(o\right)}$ represents the flux-normalized propagating wave amplitude for an input (output) at channel $E_{n}$. The evanescent 
channels correspond to $\mathrm{Im}k_{n}>0$ where the input (output) amplitudes $\alpha_{n}^{i\left(o\right)}$ denote the forward 
evanescent waves.}

\tp{
The Floquet $S$ matrix is defined as
\begin{equation}
\ensuremath{\varPsi^{o}=\mathcal{S}\varPsi^{i}},\label{eq: def_S}
\end{equation}
where the input/output amplitudes are 
\begin{align}
\left(\varPsi^{i/o}\right)^{T}\equiv & \left(\cdots,L_{n-1}^{i/o},R_{n-1}^{i/o},L_{n}^{i/o},R_{n}^{i/o},L_{n+1}^{i/o},R_{n+1}^{i/o},\cdots\right).
\label{eq: I_o_states}
\end{align}
The difference (appearence of phase) in the definition of the Floquet $S$ matrix in Eq.~(\ref{eq: I_o_states}) and the one in Eq.~(\ref{eq: I_O}) 
is artificial and it is due to different labeling of the sites in each of these approaches. We proceed with the derivation of the Floquet $S$ matrix. 
First we construct the stationary scattering states excited from each possible single-channel input $\alpha_{n}^{i}=1$. The corresponding scattering 
field amplitudes inside the scatterimng system is written as $\left|\psi_{\left(\alpha_{n}\right)}^{S}\right\rangle $, where each entry $\Braket{P|
\psi_{\left(\alpha_{n}\right)}^{S}}$ denotes the field amplitude at the mode $P$. These field amplitudes satisfy the coupled-mode equations}
\tp{
\begin{align}
\imath\frac{\mathrm{d}}{\mathrm{dt}}\Braket{P|\psi_{\left(\alpha_{n}\right)}^{S}}=\sum_{P'} & H_{PP'}\left(t\right)\Braket{P'|\psi_{\left(\alpha_{n}\right)}^{S}}+\sum_{\beta}\tilde{W}_{P\beta}\psi_{\left(\alpha_{n}\right)}^{\beta}\left(-1\right),\label{eq: EOM}
\end{align}
where $\tilde{W}_{P\beta}=-1$ when the mode $P$ is connected directly with the lead $\beta$ and $\tilde{W}_{P\beta}=0$ otherwise. Likewise,
in the case of the single-channel input $\alpha_{n}^{i}=1$, the field amplitude at site $x_{\beta}=-1$ (associated with the lead $\beta$) is denoted 
as $\psi_{\left(\alpha_{n}\right)}^{\beta}\left(-1\right)$. Using Eqs.~(\ref{eq: ansatz},\ref{eq: def_S}), we get the following expression
for the field amplitude $\psi_{\left(\alpha_{n}\right)}^{\beta}\left(-1\right)$:
\begin{align}
\psi_{\left(\alpha_{n}\right)}^{\beta}\left(-1\right)= & \frac{\delta_{\beta\alpha}}{\sqrt{\sin k_{n}}}e^{-\imath E_{n}t+\imath k_{n}\left(-1\right)}+
\sum_{m}\frac{S_{\beta_{m}\alpha_{n}}}{\sqrt{\sin k_{m}}}e^{-\imath E_{m}t-\imath k_{m}\left(-1\right)}.\label{eq: m1_term}
\end{align}
Next we substitute the Fourier series $H\left(t\right)=\sum_{n}H^{\left(n\right)}e^{-\imath n\omega t}$ together with the ansatz 
$\left|\psi_{\left(\alpha_{n}\right)}^{S}\right\rangle =\sum_{m}\left|\psi_{\left(\alpha_{n}\right)}^{S,\left(m\right)}\right\rangle e^{-\imath E_{m}t}$
and with Eq.~(\ref{eq: m1_term}) into Eq.~(\ref{eq: EOM}). Matching together each harmonic term $\propto e^{-\imath E_{m}t}$ leads to}
\tp{
\begin{align}
\sum_{P'}\sum_{m'}\left[E_{m}\delta_{PP'}\delta_{m,m'}-H_{PP'}^{\left(m-m'\right)}\right]\Braket{P'|\psi_{\left(\alpha_{n}\right)}^{S,\left(m'\right)}} 
=\sum_{\beta}\tilde{W}_{P\beta}\left(\frac{\delta_{\beta\alpha}\delta_{mn}}{\sqrt{\sin k_{n}}}e^{-\imath k_{n}}+\frac{S_{\beta_{m}\alpha_{n}}}
{\sqrt{\sin k_{m}}}e^{\imath k_{m}}\right).
\label{eq: EOM_comp}
\end{align}
Eq.~(\ref{eq: EOM_comp}) can be rewritten in a matrix form as 
\begin{align}
\left(E_{D}-H_{Q}^{0}\right)\Psi^{S}= & \tilde{\tilde{W}}\left(K^{-}+K^{+}S\right),\label{eq: Matrix1}
\end{align}
where the matrix components $\left(E_{D}\right)_{P_{m},P'_{m'}}=E_{m}\delta_{PP'}\delta_{m,m'}$,
$\left(H_{Q}^{0}\right)_{P_{m},P'_{m'}}=H_{PP'}^{\left(m-m'\right)}$,
$\left(\Psi^{S}\right)_{P_{m},\alpha_{n}}=\Braket{P|\psi_{\left(\alpha_{n}\right)}^{S,\left(m\right)}}$,
$\left(\tilde{\tilde{W}}\right)_{P_{m},\alpha_{n}}=\tilde{W}_{P\alpha}\delta_{mn}$
and $\left(K^{\pm}\right)_{\beta_{m},\alpha_{n}}=\frac{\delta_{\beta\alpha}\delta_{mn}}{\sqrt{\sin k_{n}}}e^{\pm\imath k_{n}}$.}

\tp{At the same time, each lead $\beta$ is connected with one mode of the system at the site $x_{\beta}=0$. Therefore we have $\psi_{\left(\alpha_{n}\right)}^{\beta}\left(0\right)=-\sum_{P}\left(\tilde{W}^{T}\right)_{\beta P}\Braket{P|\psi_{\left(\alpha_{n}\right)}^{S}}$.
After matching each harmonic term $\propto e^{-\imath E_{m}t}$ we get
\begin{align}
\frac{\delta_{\beta\alpha}\delta_{mn}}{\sqrt{\sin k_{m}}}+\frac{S_{\beta_{m},\alpha_{n}}}{\sqrt{\sin k_{m}}}= & -\sum_{P}\sum_{m'}\left(\tilde{W}^{T}\right)_{\beta P}\delta_{mm'}\Braket{P|\psi_{\left(\alpha_{n}\right)}^{S,\left(m'\right)}},\label{eq: EOM2_comp}
\end{align}
which can be also written in a matrix form as}
\tp{
\begin{align}
N_{k}\left(I+S\right)= & -\tilde{\tilde{W}}^{T}\Psi^{S}.
\label{eq: Matrix2}
\end{align}
with $\left(N_{k}\right)_{\beta_{m}\alpha_{n}}=\frac{\delta_{\beta\alpha}\delta_{mn}}{\sqrt{\sin k_{m}}}$. }
\tp{
Combining Eqs.~(\ref{eq: Matrix1}) and (\ref{eq: Matrix2}) allows us to eliminate the matrix $\Psi^{S}$, and eventually express the $S$-matrix as 
\begin{align}
S= & -1+2\imath W^{T}\frac{1}{D_{M}-H_{Q}^{0}+\imath WW^{T}}W\label{eq: Floquet_S}
\end{align}
where the coupling matrix $W\equiv\tilde{\tilde{W}}N_{k}^{-1}$ and
$D_{M}\equiv E_{D}+\tilde{\tilde{W}}C_{k}\tilde{\tilde{W}}^{T}$ with
$\left(C_{k}\right)_{\beta_{m}\alpha_{n}}=\delta_{\beta\alpha}\delta_{mn}\cos k_{m}$.
Note that in this derivation, we have used the Sherman-Morrison-Woodbury formula
for the matrix inverse of sum\cite{Meyer}. }

\tp{In the case of a two-mode system, Eq.~(\ref{eq: Floquet_S}) can be further simplified to the form shown in the main text. It reads 
\begin{align}
S= & -1+2\imath W^{T}\frac{1}{-D-H_{Q}^{0}}W,\label{eq: S_two_level}
\end{align}
where $\left(D\right)_{P_{m},P'_{m'}}=\delta_{PP'}\delta_{mm'}e^{-\imath k_{m}}$
and we have used the relation $E_{n}=-2\cos k_{n}$.}

\tp{It is important to point out that the NRT studied in this paper is understood non-stroboscopically. Indeed, based on the stationary solution 
Eq.~(\ref{eq: ansatz}) of the Floquet driving problem, we can see that an arbitrary change of the time origin, say $t\rightarrow t+t_{0}$, could 
give only different phases to the input/output amplitudes for different channels $E_{n}$, which affect the entry phase of the Floquet $S$ matrix, 
see Eqs.~(\ref{eq: def_S}) and (\ref{eq: I_o_states}). In this respect, the non-reciprocity parameter defined in Eq. (6) of the main text remains 
unchanged, because it involves absolute values of the $S$ matrix entries. This point can be further appreciated via a further analysis of
Eq.~(\ref{eq: Floquet_S}) or Eq.~(\ref{eq: S_two_level}) for the Floquet $S$ matrix. Consider for simplicity Eq.~(\ref{eq: S_two_level}) associated
with a two-mode system. As $t\rightarrow t+t_{0}$ we have that the time-dependent Hamiltonian $H\left(t\right)\rightarrow H_{new}\left(t\right)=
H\left(t+t_{0}\right)$, and thus $H_{Q}^{0}$ in Eq.~(\ref{eq: S_two_level}) becomes $H_{new,Q}^{0}=\mathcal{T}_{0}^{-1}H_{Q}^{0}\mathcal{T}_{0}$,
where $\left(\mathcal{T}_{0}\right)_{P_{m},P'_{m'}}=\delta_{PP'}\delta_{mm'}e^{\imath m\omega t_{0}}$. Consequently, we have from 
Eq.~(\ref{eq: S_two_level}) that $S_{new}=\mathcal{T}_{0}^{-1}S\mathcal{T}_{0}$, which confirms the invariance of the non-reciprocity parameter 
$NR_{n\rightarrow m}$. It is worth pointing out that the change of time origin could affect the ``nature'' of bonds, say from directional to directionless or
vice versa. But since the change of the nature of all the bonds is ``coherent", the interference features between paths of different directionality
are not affected. Thus NRT remains the same. }

\section{Floquet Design for Example 3}

We explain the design of the Floquet modulation, associated with a {\it real} driving of the coupling $h\left(t\right)$ between the 
two resonators, that leads to nonreciprocal transport $NR_{0\rightarrow0}\neq 0$. The design strategy consists of three steps: 
First we shall analyze the connectivity of the Floquet ladder (Floquet lattice) associated with such real-valued driving. Then we 
shall impose the two necessary conditions for non-reciprocal transport that we have discussed in the paper and identify the 
constrains that they impose on the topology/connectivity of the Floquet lattice. Finally using the grammar shown in Fig.~2 of the 
paper we shall identify the existing bonds with corresponding driving Fourier components. This will allow us to ``translate" the 
designed Floquet lattice to a driven Hamiltonian in the actual space and identify the functional form of the coupling driving 
scheme $h(t)$ that is associated with the Floquet construction.

We start with some general observations about the Floquet lattice associated with the real-valued coupling driving schemes.
A direct consequence of the grammar (see Fig. 2) is that the Floquet bonds satisfy a mirror symmetry with respect to the bisector 
between the L and R Floquet sites. Moreover, the coupling strengths associated with the opposite 
directions of each bond are complex conjugate of one-another. Finally, the vertical bonds that connect only L (R) with L (R) 
sites in the Floquet lattice vanishes since the resonant frequency (on-site potential) of the resonators is not modulated in
time, see Fig.~2.

In the scattering framework, the nonreciprocity $NR_{0\rightarrow0}$ is associated with the difference between the absolute-valued 
Green functions $|G\left(0,R;0,L\right)|$ and $|G\left(0,L;0,R\right)|$. Both of them can be expressed in terms of an infinite sum of 
{\it directed} paths which connect the sites $\left(0,L\right)$ and $\left(0,R\right)$ in the Floquet lattice. It is important to stress
that while the set of the infinite paths appearing in these sums is the same, their transport direction is opposite in each of these 
cases. When the absolute value is considered various interferences between these directed paths can lead to differences between 
$|G\left(0,R;0,L\right)|$ and $|G\left(0,L;0,R\right)|$.

The first necessary condition for the existence of non-reciprocal transport is the presence of a directed path i.e. a path that has 
different transition amplitudes (by a phase) for forward and backward propagation. This imposes a stronger constrain than the 
requirement that the constituent bonds (of the path) are directed. Let us consider, for example, the path shown in Fig.~S2a. In 
this figure any complex-valued bond strength is indicated with a tilde e.g. ${\tilde h}_1$, while real-valued bond strengths are 
indicated without tilde e.g. $h_0$.
It is easy to show that its contribution to the Green functions $G\left(0,R;0,L\right)$ and $G\left(0,L;0,R\right)$ is proportional to $h_{0}
\tilde{h}_{1}\tilde{h}_{1}^{*}$, which lacks directionality even though the constituent bonds are directed by themselves (the suppressed 
common proportionality constants are simply the multiplications of the locators $g_{n}^{L/R}$, which are irrelevant for the directionality). 
In fact, one can further show that, within the nearest-neighbor couplings of the Floquet ladder, all paths that connect the site $(0,L)$ to 
the site $(0,R)$ are lacking directionality. 

Therefore, a nonreciprocal transport requires the design of a Floquet lattice which has (at least!) bonds that couple next-nearest-
neighbor sites on the Floquet ladder. Indeed, such ladder can support directional paths, see for example Fig.~S2b. Specifically, the 
contribution of this path to the Green functions $G\left(0,R;0,L\right) (G\left(0,L;0,R\right))$ is  $\propto\tilde{h}_{1}\tilde{h}_{2}^{*}
\tilde{h}_{1} (\propto\tilde{h}_{1}^{*}\tilde{h}_{2}\tilde{h}_{1}^{*})$. Below, for simplicity, we assume that $\tilde{h}_{j}=-\imath h_{j},j=1,2$
and $h_{j}$ are real numbers, \tp{which are consistent with the grammar in Fig.~2 of the main text. }In this case $\tilde{h}_{1}\tilde{h}_{2}^{*}\tilde{h}_{1}=- \tilde{h}_{1}^{*}\tilde{h}_{2}\tilde{h}_{1}^{*}$.

The second necessary condition for nonreciprocal transport is the existence of direction-dependent interferences between two paths. 
This condition can be easily realized by introducing a horizontal coupling $h_{0}$ to the Floquet lattice. The interference of these two 
paths is shown in Fig.~S2c. Approximating the associated Green functions as a sum of these two paths results to the following two 
expressions

\begin{align}
\label{S1}
|G\left(0,R;0,L\right)|\approx & |g_{0}^{L}g_{0}^{R}|\left|\left\{ h_{0}+\left(-\imath h_{1}\right)g_{-1}^{R}\left(\imath h_{2}\right)g_{1}^{L}\left(-\imath h_{1}\right)\right\} \right|
\end{align}
and 
\begin{align}
\label{S2}
|G\left(0,L;0,R\right)|\approx & |g_{0}^{L}g_{0}^{R}|\left|\left\{ h_{0}+\left(\imath h_{1}\right)g_{1}^{L}\left(-\imath h_{2}\right)g_{-1}^{R}\left(\imath h_{1}\right)\right\}\right| .
\end{align}
Direct inspection of Eqs. (\ref{S1},\ref{S2}) indicates that if ${\cal I}m\left(g_{n}^{L/R}\right)=0$ then the Green functions $G\left(0,R;0,L\right)$ 
and $G\left(0,L;0,R\right)$ are complex conjugates of each other and therefore have the same absolute value. Obviously in this case 
$NR_{0\rightarrow0}=0$. This observation leads us to the final step associated with the construction of the Floquet lattice i.e. the locators $g_{n}^{L/R}$
must have a nonzero imaginary part. The nonzero imaginary part of the locators $g_{n}^{L/R}$ could be due to the material loss/gain
or other leaking propagating channels for small driving frequency,
say $E_{\pm1}$ in this case.

\begin{figure}[h]
\includegraphics[width=0.9\columnwidth,keepaspectratio,clip]{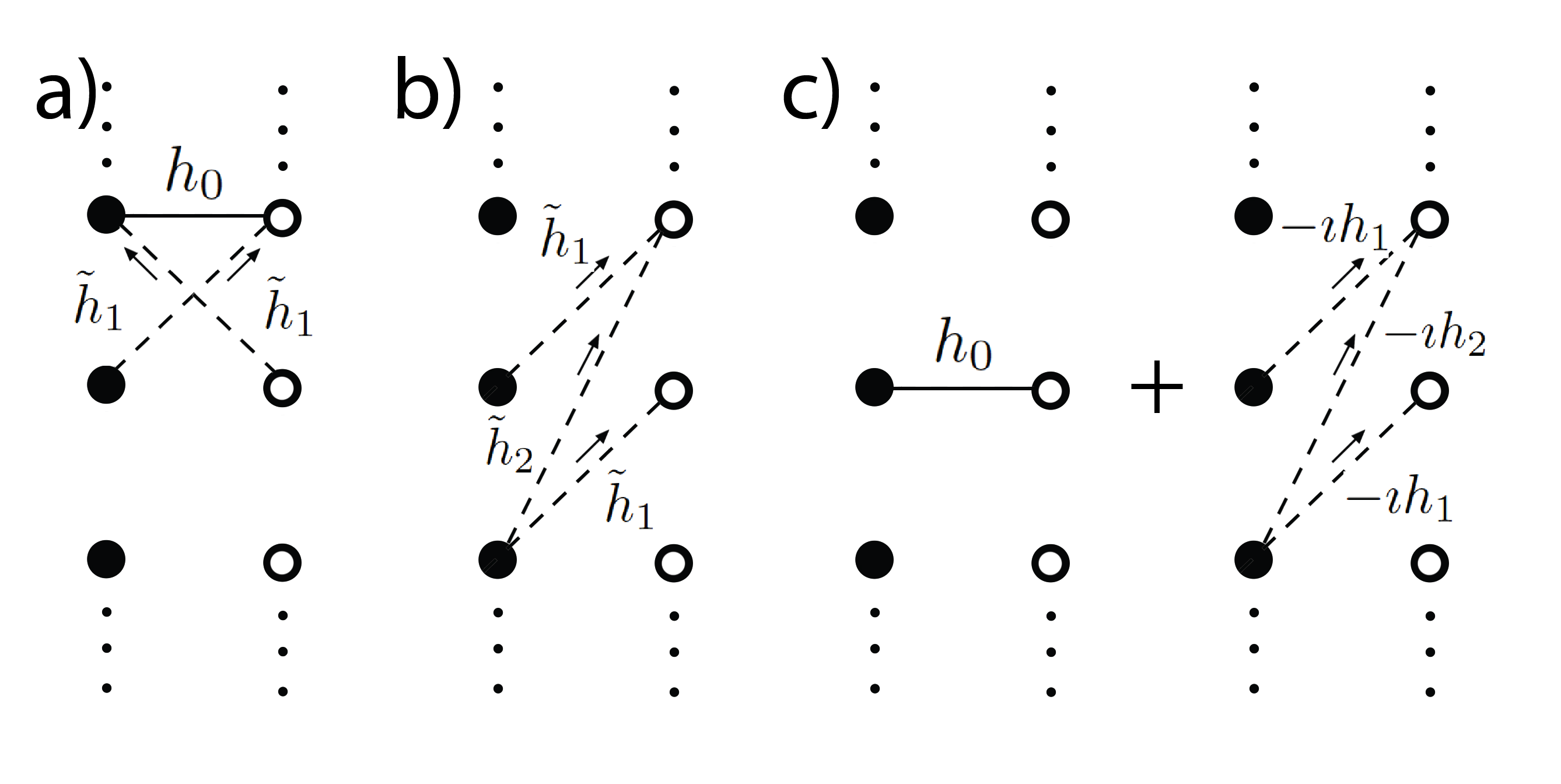}

\caption{ (color online) a) Transport-direction independent path  with directed bonds and b) transport-direction dependent path. c) The direction-dependent  interference between two  paths, which is responsible for the nonreciprocity  transport. 
}
\label{Sfig2}
\end{figure}

Finally we transform the designed Floquet lattice to the equivalent driving scheme. This can be done easily by translating the
Floquet  bonds of Fig. S2c to the associated driving matrices using the grammar shown in Fig. 2. Indicating the strength of the non-directed
horizontal coupling  as $h_{0}$ and the directed diagonal couplings up to the next-nearest-neighbor with upward transition amplitudes as 
$-\imath h_{1}$ and $-\imath h_{2}$ we get $h\left(t\right)=h_{0}+2h_{1}\sin\omega t+2h_{2}\sin2\omega t$.

\begin{figure}[h]
\includegraphics[width=0.8\columnwidth,keepaspectratio,clip]{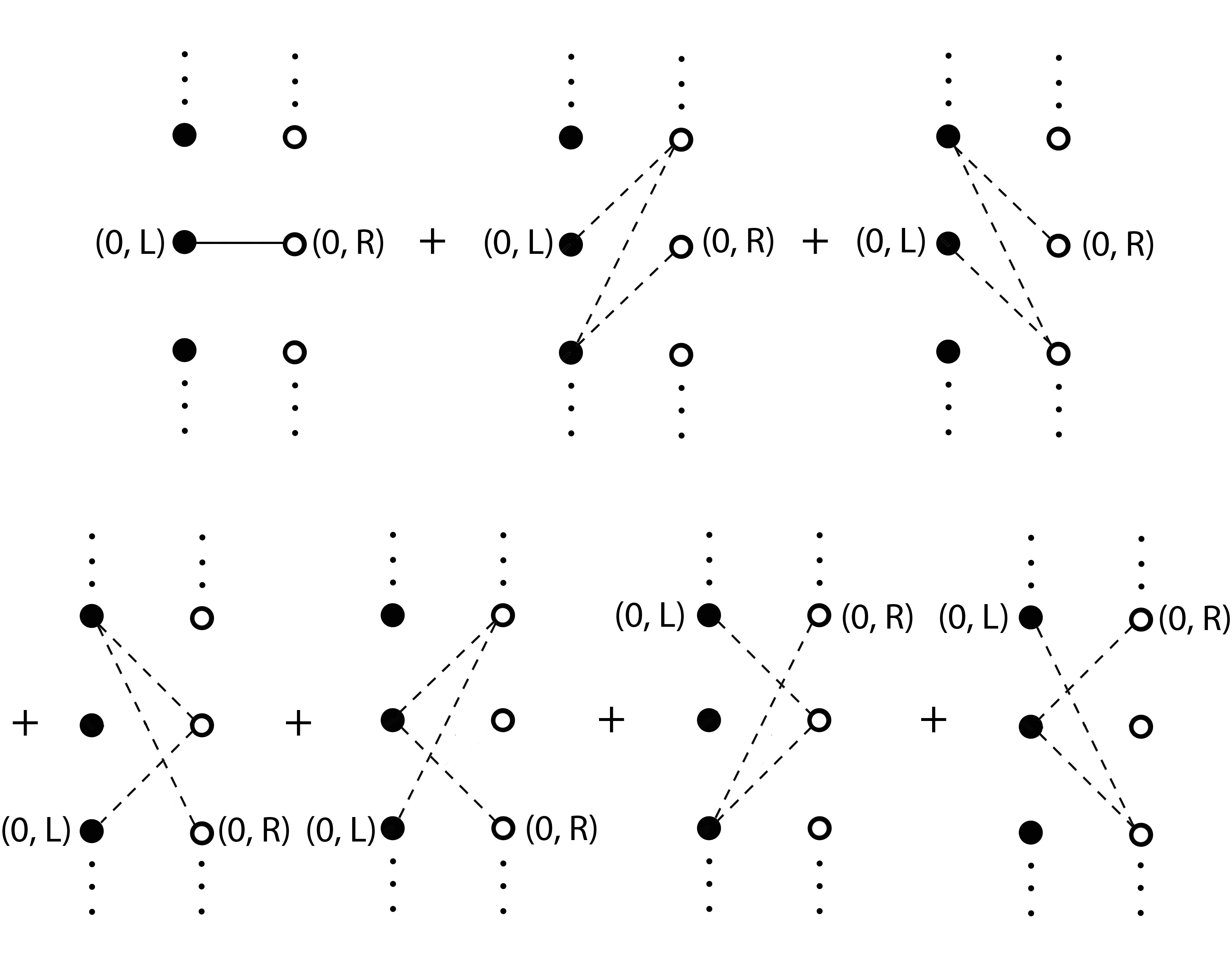}

\caption{ (color online) The interference paths used to obtain approximated non-reciprocity strength $NR_{0\rightarrow0}$ in Eq.~\ref{eq: NR3}.
}
\label{Sfig3}
\end{figure}

\tp{
For the theoretical evaluation of $NR_{0\rightarrow0}$, we consider the paths shown in Fig.~\ref{Sfig3}. Notice that in this case diagrams similar to the ones shown in Fig. \ref{Sfig2}a do not contribute to this approximation. These diagrams, although they involve directional bonds (see dashed lines), are altogether directionless as explained above. Therefore they are considered of higher order as compared to the basic directionless path involving an $h_0$ bond (see first diagram in Fig. \ref{Sfig3}.)
Correspondingly, the Green's functions are $G\left(0,R;0,L\right)=g_{0}^{L}g_{0}^{R}\left(h_{0}-\imath h_{1}^{2}h_{2}\Delta\right)$
(and $G\left(0,L;0,R\right)=g_{0}^{L}g_{0}^{R}\left(h_{0}+\imath h_{1}^{2}h_{2}\Delta\right)$),
where $\Delta=g_{-1}^{R}g_{1}^{L}-g_{-1}^{L}g_{1}^{R}+g_{-1}^{R}g_{-2}^{L}-g_{-1}^{L}g_{-2}^{R}-g_{1}^{R}g_{2}^{L}+g_{1}^{L}g_{2}^{R}$.
Thus the non-reciprocity strength $NR_{0\rightarrow0}$ is }
\tp{
\begin{align}
NR_{0\rightarrow0}\approx & \frac{2h_{0}h_{1}^{2}h_{2}\mathrm{Im}\Delta}{h_{0}^{2}+\left(h_{1}^{2}h_{2}\right)^{2}\left|\Delta\right|^{2}}.
\label{eq: NR3}
\end{align}
where for its calculation we have used Eqs.~(2, 6) of the main text.  }

\begin{figure}[h]
\includegraphics[width=0.8\columnwidth,keepaspectratio,clip]{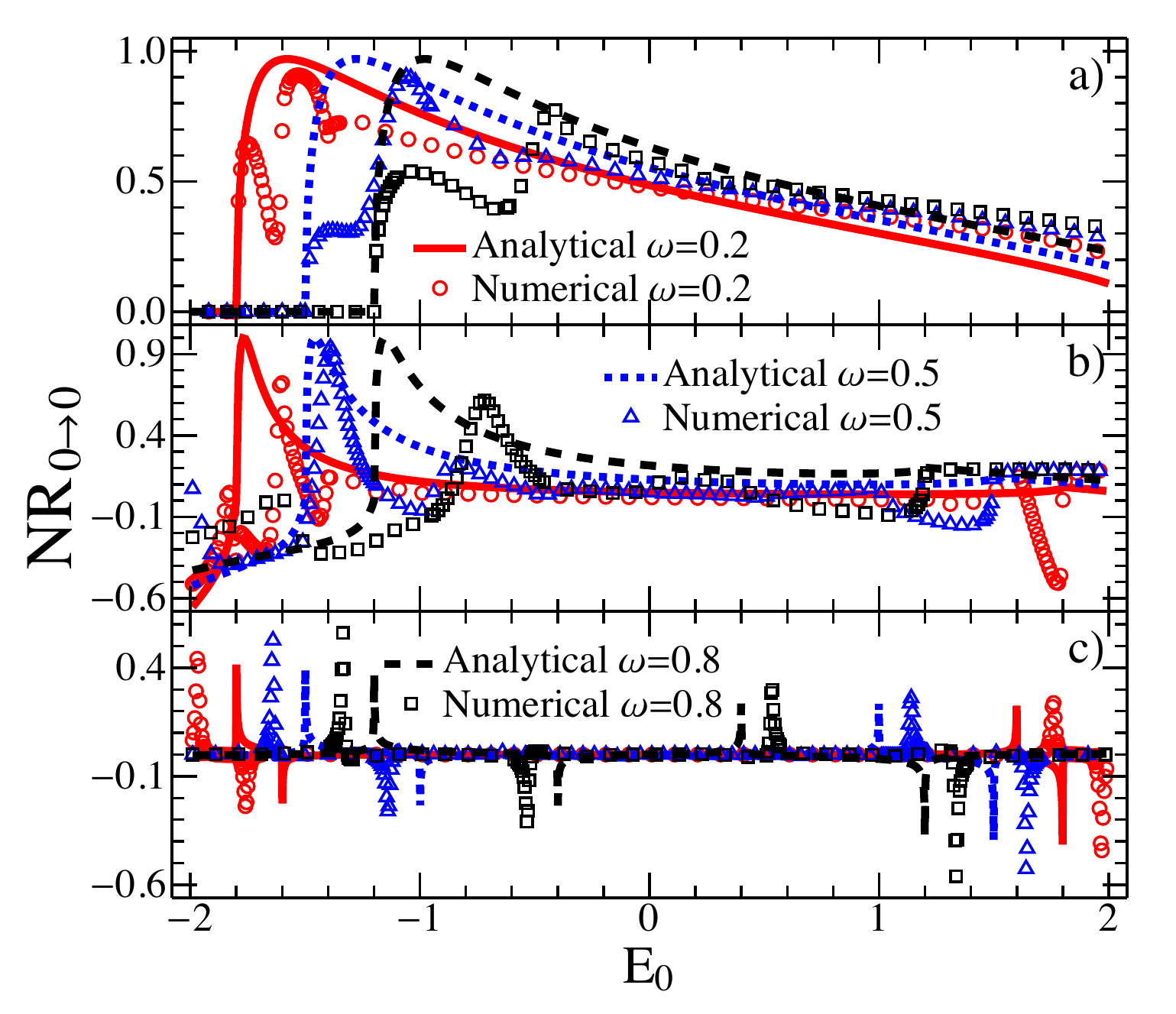}
\caption{ (color online) Numerical (symbols) and theoretical (lines) results of $NR_{0\rightarrow0}$ versus incident frequency $E_0$ 
when driving frequencies $\omega=0.2,0.5,0.8$ for a) Example 1 with other parameters being $\varepsilon_{L}^{0}=-1$, $\varepsilon_{R}^{0}
=0$, $h_0=-0.5$,  $f_{1}=h_{1}=0.5$; b)Example 2 with other parameters being $\varepsilon_{L}^{0}=-1$, $\varepsilon_{R}^{0}=0$, 
$h_0=-1$,  $f_{1}=h_{1}=0.5$ and  c) Example 3 with other parameters being $\varepsilon_{L}^{0}=-1$, $\varepsilon_{R}^{0}=1$, $h_0
=-1$,  $h_{1}=h_{2}=0.2$. The  other parameters are the same as the ones used  in Fig.~3 of the main text for the case of small driving 
frequency $\omega=0.5$. 
}
\label{Sfig4}
\end{figure}

\tp{
Next we study the effect of the magnitude of the driving frequency on the non-reciprocity parameter $NR_{0\rightarrow0}\left(E_{0}\right)$. In Fig.~\ref{Sfig4}, we
show both numerical and theoretical results for each of the three examples discussed in the main text. For each driving scheme we have considered three values
of driving frequencies $\omega=0.2,0.5,0.8$. We find that, even in the case of broad-band NRT, the driving frequency controls the position of the incident energy
$E_0$ where maximum non-reciprocity is observed. This observation is better appreciated in the case of narrow-band NRT (see example 3) where non-reciprocal 
resonant peaks occur due to strong interferences between the different directional paths. The latter occurs when $\left|g_{0}^{L}g_{0}^{R}h_{0}\right|\sim \left|g_{0}^{L}
g_{0}^{R}\imath h_{1}^{2}h_{2}\Delta\right|$. In the (typical) case when $h_{1},h_{2}\ll h_{0}$, this approximate relation occurs in a narrow window 
around the resonant energies of the locators $\Delta$. It is important to note that this insight into the NRT rely heavily on the locator expansion approach 
for $G\left(0,R;0,L\right)$ (or similarly  $G\left(0,L;0,R\right)$) . }

\begin{figure}[h]
\includegraphics[width=0.8\columnwidth,keepaspectratio,clip]{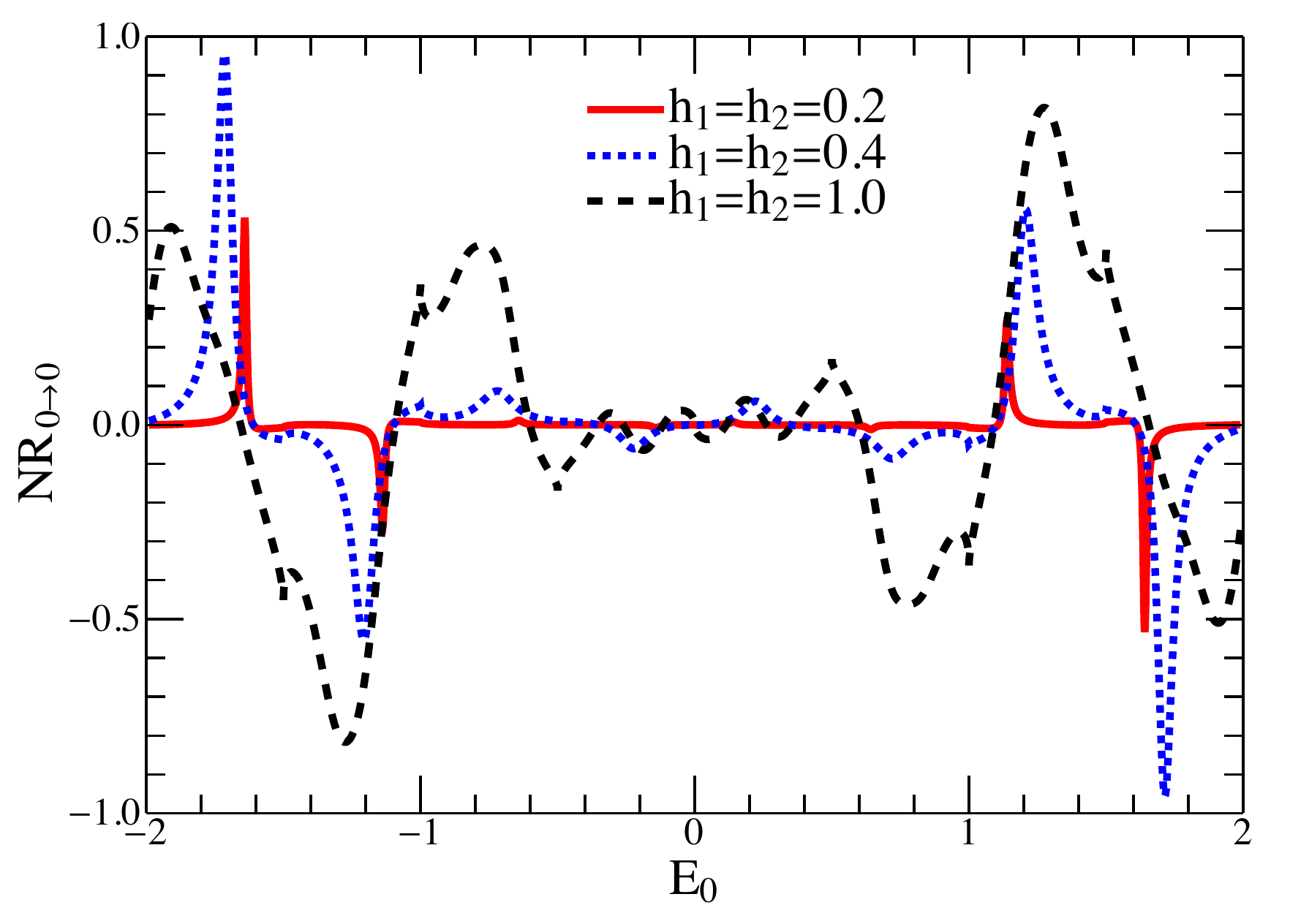}
\caption{ (color online) Numerical evaluation of $NR_{0\rightarrow 0}$ for example 3 with $h_0=-1$ and three different values of $h_1=h_2=0.2, 0.4, 1$ which demonstrate 
a control of the NRT band-width. The other parameters are $\omega=0.5$, $\epsilon_L^0=-1$ and $\epsilon_R^0=1$.
}
\label{Sfig4b}
\end{figure}

\tp{In Fig. \ref{Sfig4b} we demonstrate the NRT band-width control for example 3 by varying the driving amplitudes $h_1=h_2$ (see discussion in the main text). The 
management of the NRT bandwidth 
$\delta E$ can be achieved by choosing appropriate values of $h_1=h_2$ that violate or satisfy the relation $|{\tilde \eta}|\sim |\eta|$. In this example, we have
that $|\eta|\sim h_0$. At the same time ${\tilde \eta}\approx B*L(E_0)$ where $B=h_1^2 h_2$ and $L(E_0)=|\Delta|$. When $B=0.2^3$ (corresponding to $h_1=h_2=0.2$,
see red line in Fig. \ref{Sfig4b}), the condition $|{\tilde \eta}|\sim |\eta|$  is violated leading to very weak NRT. The only exception appears when the incident energies $E_0$ 
are close to the resonant energy of $\Delta$. At these energies (see Fig. \ref{Sfig4b}) a sharp (narrow-band) NRT peak(s) emerge. As $h_1=h_2$ increase, $B$ increases 
as well (see blue and black lines in Fig. \ref{Sfig4b}) leading to a validation of the condition $|{\tilde \eta}|\sim |\eta|$ for a broad frequency range away from the resonant 
energies. Note also that within these windows, i.e. away from resonances, $|\Delta(E_0)|$ is a smooth function of $E_0$. In this case we have a broad-band non-reciprocal 
transport.}

\tp{Let us finally discuss the source of the deviation between the numerical and the theoretical results Eq.~(\ref{eq: NR3}). For presentation 
purposes, we focus on the driving scheme associated with example 3, see Fig. \ref{Sfig4}c. To this end, we rewrite Eq. (3) of the main text 
as $G=\left(1-G^{0}V_{Q}\right)^{-1}G^{0}$. In order to evaluate the Green's function $G$ and the consequent $NR_{0\rightarrow0}$, we
now truncate both the matrices $G^{0}$ and $V_{Q}$ in the channel space $\left(n,\alpha\right)$ with $n=-1,0,1$ and $\alpha=L,R$. This
results in a $6\times6$ finite matrices for $G^{0}$ and $V_{Q}$ while the only locators $g_{n}^{\alpha}$ that appear in the expansion correspond
to $n=0,\pm1$. Under this approximation, and taking into account that $h_{1}, h_{2}\ll h_{0}$, we get $\widetilde{G}\left(0,R;0,L\right)\propto 
h_{0}N_{r}-\imath h_{1}^{2}h_{2}\tilde{\Delta}$ ($\widetilde{G}\left(0,L;0,R\right)\propto h_{0}N_{r}+\imath h_{1}^{2}h_{2}\tilde{\Delta}$). 
The renormalization parameter $N_r$ for the directionless path is $N_{r}=1-h_{2}^{2}g_{1}^{L}g_{-1}^{R}-h_{2}^{2}g_{-1}^{L}g_{1}^{R}-h_{0}^{2}
\left(g_{-1}^{L}g_{-1}^{R}+g_{1}^{L}g_{1}^{R}\right)+h_{0}^{4}g_{1}^{L}g_{-1}^{L}g_{1}^{R}g_{-1}^{R}$,  $\tilde{\Delta}=
g_{-1}^{R}g_{1}^{L}-g_{-1}^{L}g_{1}^{R}$ while the proportionality constants for both ${\tilde G}\left(0,R;0,L\right)$ and
${\tilde G}\left(0,L;0,R\right)$ are the same. Substituting all the above in the expression for the $\widetilde{NR}_{0\rightarrow0}$ we get
\begin{align}
\widetilde{NR}_{0\rightarrow0}\approx & \frac{2h_{0}h_{1}^{2}h_{2}\left({\cal I}m\tilde{\Delta} {\cal R}e
N_{r}-{\cal I}mN_{r}{\cal R}e\tilde{\Delta}\right)}{h_{0}^{2}\left|N_{r}\right|^{2}+\left(h_{1}^{2}h_{2}\right)^{2}
\left|\tilde{\Delta}\right|^{2}}.
\label{eq: NR3_norm}
\end{align}}
\tp{where we have used {\it tilde} in order to indicate that the $NR$ is calculated using the previous truncation scheme.}

\begin{figure}[h]
\includegraphics[width=0.8\columnwidth,keepaspectratio,clip]{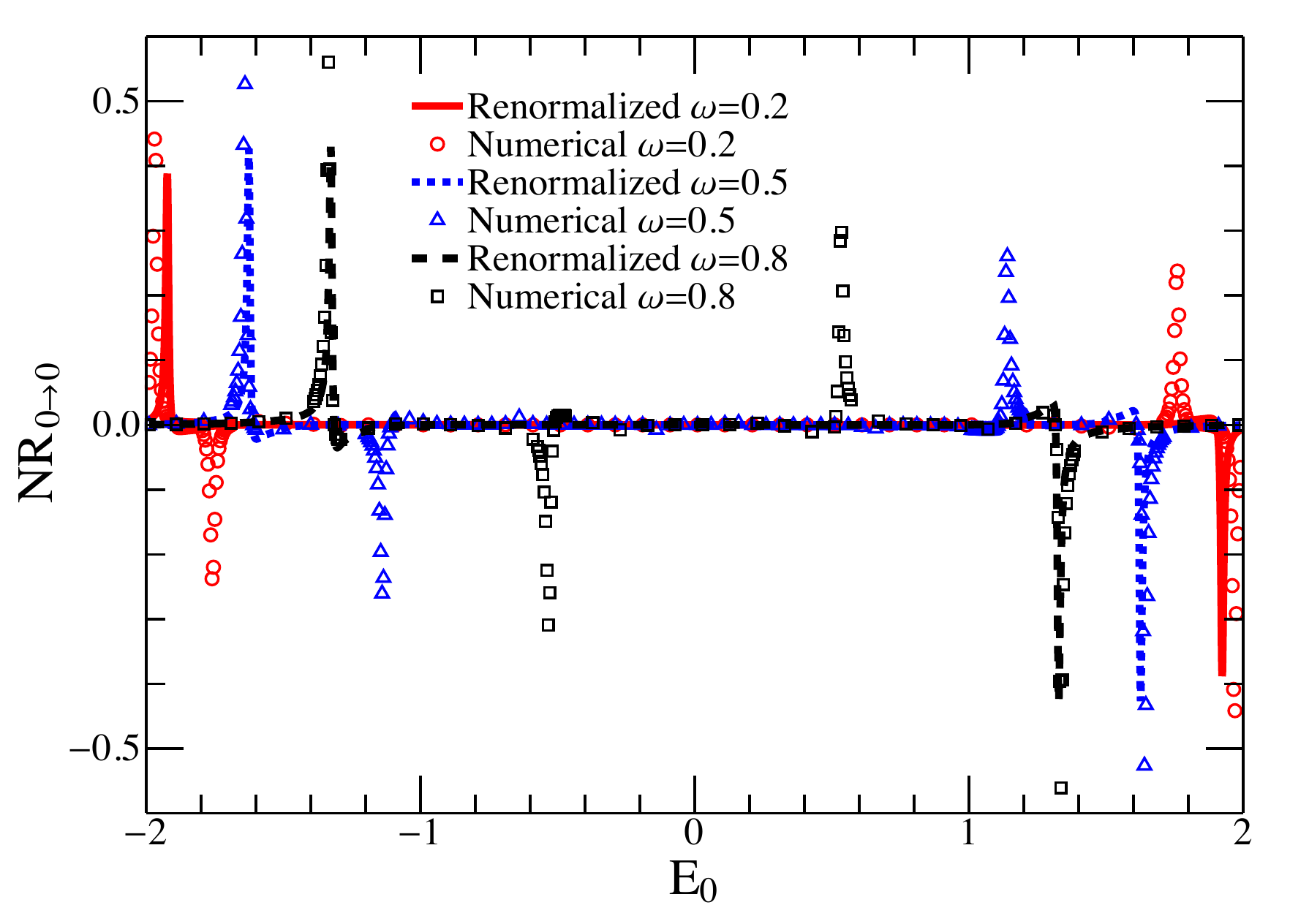}
\caption{ (color online) Numerical (symbols) and renormalized theoretical Eq.~(\ref{eq: NR3_norm}) (lines) results of 
$NR_{0\rightarrow0}$ versus incident frequency $E_0$  when driving frequencies $\omega=0.2,0.5,0.8$ for  example 
3 with the same other parameters as  Fig.~\ref{Sfig4}.
}
\label{Sfig5}
\end{figure}

\tp{The comparison of Eq.~(\ref{eq: NR3_norm}) (lines) with the corresponding numerical results (symbols) is shown in Fig.~\ref{Sfig5}. 
We found that Eq.~(\ref{eq: NR3_norm}) captures nicely the NRT resonant peak close to the left/right edges of the energy band. We 
conclude therefore that the deviations between the numerics and the theoretical results Eq.~(\ref{eq: NR3}) is due to the renormalization 
factor $N_{r}$ which is present in the expression Eq. (\ref{eq: NR3_norm}). Let us finally comment that the NRT resonant peaks near the 
center of the band are not captured by our calculations Eq. (\ref{eq: NR3_norm}), and they require to consider higher-order locators 
(especially $g_{2}^{\alpha}$, see Eq.~(\ref{eq: NR3})). Instead, these peaks are captured (to some extent) by Eq. (\ref{eq: NR3}) which
involves higher-order locator $g_{2}^{\alpha}$. This allow us to conclude that different peaks originate from locator resonances 
of different order. }

\begin{figure}[h]
\includegraphics[width=0.8\columnwidth,keepaspectratio,clip]{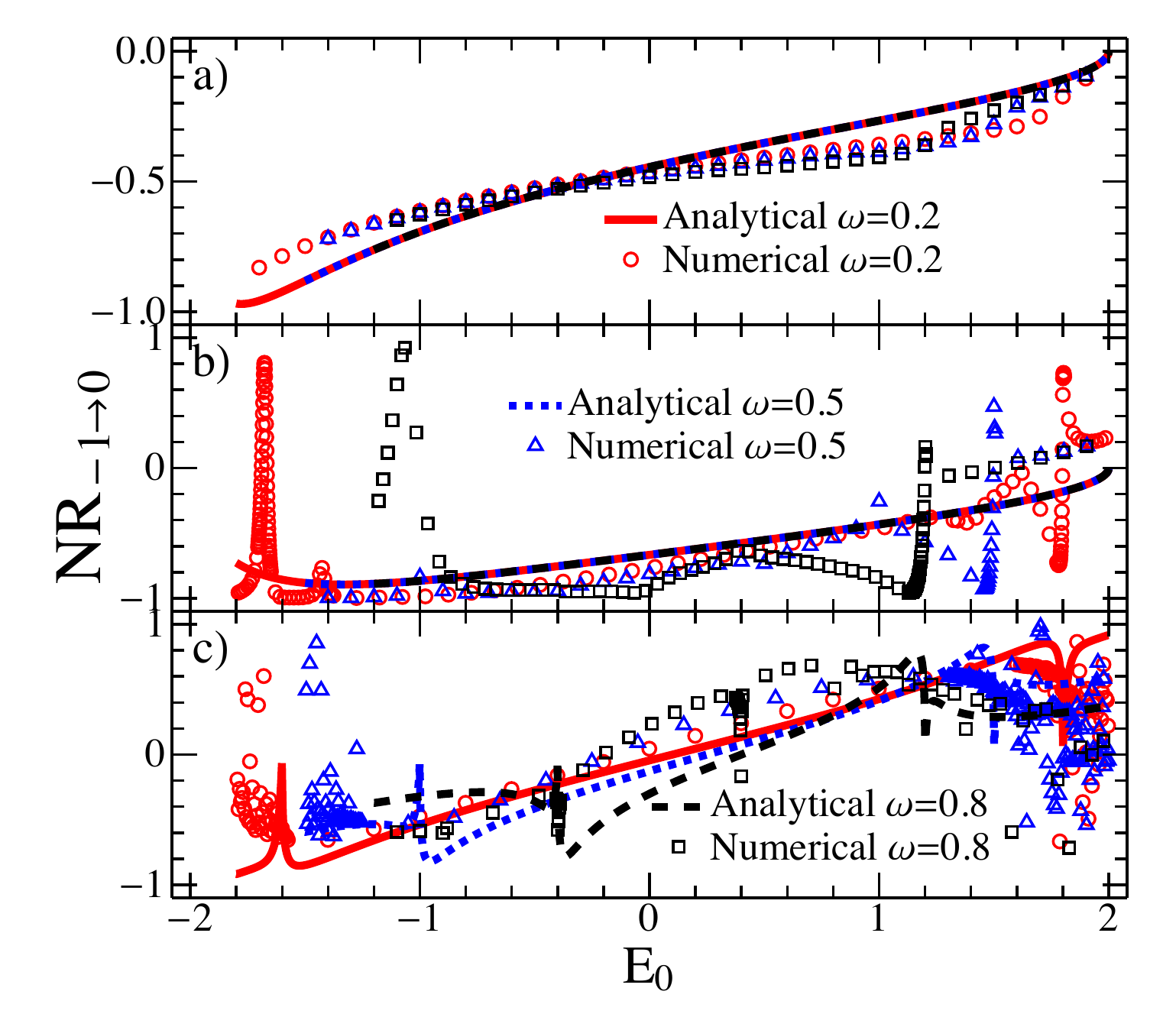}
\caption{ (color online) Numerical (symbols) and theoretical (lines) results of $NR_{-1\rightarrow0}$ versus incident energy $E_0$  for driving 
frequencies $\omega=0.2,0.5,0.8$ and a) Example 1; b)Example 2  c) Example 3. All other parameters are the same as the corresponding ones of Fig.~\ref{Sfig4}. }
\label{Sfig6}
\end{figure}

\begin{figure}[h]
\includegraphics[width=0.8\columnwidth,keepaspectratio,clip]{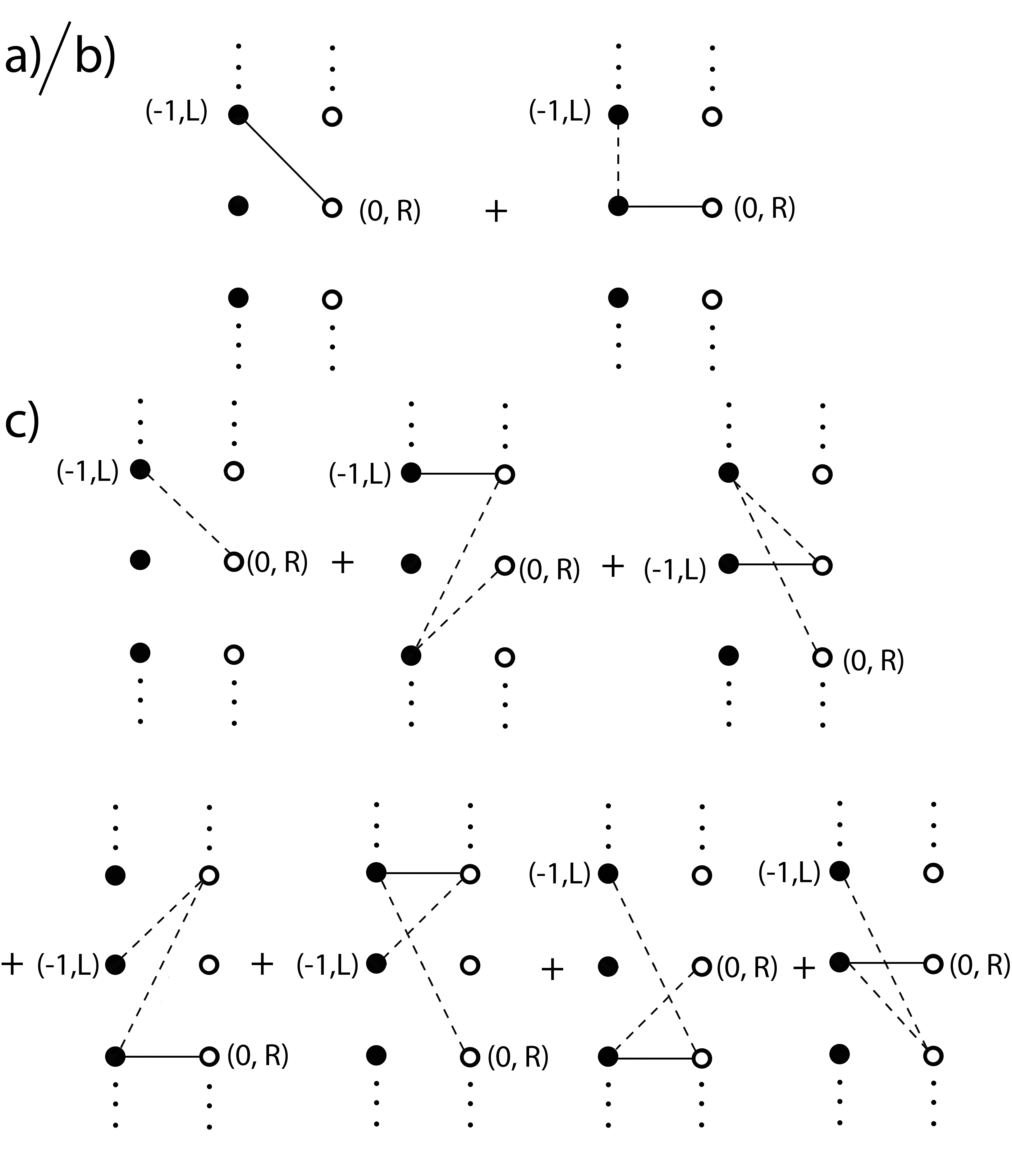}

\caption{ (color online) The interference paths used to obtain approximated non-reciprocity strength $NR_{-1\rightarrow0}$ for a)/b) Eq.~\ref{eq: up_down_12} 
of examples 1 and 2; c) Eq.~\ref{eq: up_down_3} of examples 3.
}
\label{Sfig7}
\end{figure}

\tp{
\section{Nonreciprocal up/down frequency conversion}}
\tp{Let us now consider the non-reciprocity between different energy channels $NR_{-1\rightarrow0}$ (non-reciprocal up/down
-conversion). Fig.~\ref{Sfig6} show the comparison between the numerical and the theoretical results for all three driving examples 
considered in the main text. We consider three different driving frequencies $\omega=0.2,0.5,0.8$. The paths that have been 
considered in the theoretical calculation are shown in  Fig.~\ref{Sfig7}. Notice that we use the same paths for both examples $1$ 
and $2$. The theoretical calculations follow the same lines as before. For the examples 1 and 2 we get that }
\tp{
\begin{align}
NR_{-1\rightarrow0}\approx & -\frac{2h_{0}f_{1}h_{1}\mathrm{Im}\Delta}{h_{1}^{2}+h_{0}^{2}f_{1}^{2}\left|\Delta \right|^{2}}, 
\label{eq: up_down_12}
\end{align}
where $\Delta=g_{0}^{L}$. Notice that Eq. (\ref{eq: up_down_12}) does not contain the driving frequency $\omega$ and thus we expect
that the NR up/down-conversion is insensitive to $\omega$. This prediction is also confirmed by our numerical data, see Fig. \ref{Sfig6}.
For the example 3 we have that
\begin{align}
NR_{-1\rightarrow0}\approx & \frac{2h_{0}h_{2}\mathrm{Im}\Delta}{1+h_{0}^{2}h_{2}^{2}\left|\Delta\right|^{2}}
\label{eq: up_down_3}
\end{align} }
\tp{where $\Delta=g_{-1}^{R}g_{1}^{L}+g_{-1}^{R}g_{-2}^{L}+g_{-2}^{R}g_{0}^{L}+g_{-2}^{R}g_{-2}^{L}+g_{1}^{R}g_{1}^{L}+g_{1}^{R}g_{0}^{L}$.
In this case Eq. (\ref{eq: up_down_3}) depends on the driving frequency $\omega$ via $\Delta$. However, this dependence is rather weak
as it is indicated by Fig.~\ref{Sfig6}c }

\begin{thebibliography}{99}
\bibitem{note0}Our set-up does not involve loop-configurations in physical space as opposed to Refs. \cite{ESSA14,SCA13,FSSHA14,
FYF12,KA14,LEFF14}.
\bibitem{note1} \tp{Note that $\Phi^{+}\equiv \phi(n,L;m,R)+\phi(m,R;n,L)\neq 0$ -- see the discussion of necessary conditions for NRT. Typically $\Phi^{+}(E_0)$
change slowly when $E_0$ is away from the energy corresponding to the (local) maximum of NRT. Thus it is not the main source for the band-width control 
of NRT.
}
\end{thebibliography}
\end{document}